\pdfoutput=1
\documentclass[11pt,a4paper]{article}
\usepackage{jheppub}
\newcommand{\sn}{\operatorname{sn}}
\newcommand{\cn}{\operatorname{cn}}
\newcommand{\dn}{\operatorname{dn}}
\newcommand{\EllF}{\mathrm{F}}
\newcommand{\EllE}{\mathrm{E}}
\newcommand{\EllK}{\mathrm{K}}
\newcommand{\tabbox}[2]{\par\medskip\noindent
  \fbox{\parbox{\dimexpr\linewidth-2\fboxsep-2\fboxrule\relax}{%
  \small\textbf{From the tables: #1}\par\smallskip #2}}\par\medskip}

\title{The Nambu--Jona-Lasinio model at isospin density in closed form: what a regularization can change}
\author{T. G. Khunjua}
\affiliation{The University of Georgia, GE-0171 Tbilisi, Georgia}
\emailAdd{t.khunjua@ug.edu.ge}
\abstract{At zero temperature, the mean-field thermodynamic potential of the
two-flavour Nambu--Jona-Lasinio model at isospin chemical potential reduces to a
divergent momentum integral of two quasiparticle levels.  In practice this
integral is cut off at momenta not far above the chemical potentials being
probed.  We evaluate it exactly, at an arbitrary cutoff, using the Hermite
reduction.  The result is an algebraic boundary term and three incomplete
elliptic integrals, one of each kind.  Their modulus, parameter and amplitude at
the end of the path are fixed by the triangle formed by the two levels at zero
momentum.  The power divergences reside in the boundary term, while the
remaining divergence is logarithmic: the momentum axis is an open path on an
elliptic curve, and the ultraviolet is the puncture at which it ends.

The closed form makes the role of regularization explicit.  As the cutoff is
removed, sharp and smooth cutoffs acting on the momentum alone differ only by an
additive constant and by two terms that the vacuum inputs fix.  The medium
separation scheme reproduces the resulting potential up to an additive
constant.  By contrast, a cutoff applied separately to the two levels generates
an isospin density below the onset of condensation and violates the Silver
Blaze property.

The closed form gives the state that continues the vacuum along the isospin
axis without numerical momentum integration.  With the same vacuum fit, a cutoff
kept finite gives a visibly different state.  The two differ by explicit terms
that vanish only as the cutoff is removed: they are different models, not two
regularizations of one.}

\begin{document}
\maketitle
\flushbottom
\section{Introduction}\label{sec:intro}

Strongly interacting matter, wherever nature produces it in bulk --- in a
neutron star or in the fireball of a heavy-ion collision --- carries an isospin
density along with its baryon density~\cite{BrandtEndrodiSchmalzbauer2018}.  In
both cases the isospin chemical potential is much smaller than the baryon
one~\cite{SonStephanov2001}, so that isospin density without baryon density is
an idealization.  It has nevertheless become one of the most instructive
corners of the QCD phase diagram, and the reason is a practical one.  At a
nonzero baryon density the fermion determinant is complex, and lattice Monte
Carlo fails: this is the sign problem.  At isospin density the determinant
stays real and positive~\cite{AlfordKapustinWilczek1999}, and lattice
simulations are possible~\cite{KogutSinclair2002}.  Effective
models can therefore be compared with first-principles
results~\cite{Mannarelli2019,Abbott2025}.  The picture that emerges is simple.  At zero
temperature nothing changes until $|\mu_I|$ reaches the pion mass; then charged
pions condense, and the ground state becomes a pion
superfluid~\cite{SonStephanov2001,BrandtEndrodiSchmalzbauer2018}.  Its equation
of state has even been used to build hypothetical stars made of
pions~\cite{Brandt2018PionStars}.

The Nambu--Jona-Lasinio (NJL)
model~\cite{NJL1961a,NJL1961b,Klevansky1992,Buballa2005} is one of the most
widely used effective models, and pion condensation at isospin density has been
studied in it, and in its relatives, in
detail~\cite{HeJinZhuang2005,ZhukovskyKlimenkoKhunjua2010}.  In the mean-field
approximation and at zero temperature the model comes down to a single momentum
integral.  The quasiparticles have two energy levels at each momentum, and the
thermodynamic potential consists of a term quadratic in the condensates and the
integral, over all momenta, of the sum of the two levels.  The integral
diverges.  Since the NJL model is not renormalizable, it is not defined until
one says what is to be done with this integral at large momenta.  This paper is
about that integral.  We evaluate it in closed form and use the result to
determine what a regularization can change in the potential and what it cannot.

Nothing in this integral is specific to pions.  It depends only on the
spectrum: a relativistic fermion with a mass, a chemical potential that shifts
its energy, and a gap that mixes it with a partner.  The same spectrum, and
with it the same integral, appears in other systems.  The closest is two-colour
QCD, where a baryon is made of two quarks and it is diquarks rather than pions
that condense at baryon density~\cite{Kogut2000}: the mean-field potential of the NJL model is
then ours, with the quark chemical potential in place of $\mu_I/2$ and two
colours instead of
three~\cite{KhunjuaKlimenkoZhokhov2020,Pasqualotto2026,HeMaoZhuang2013}.  Like
QCD at isospin density, two-colour QCD has no sign problem and can be simulated
on the lattice~\cite{Braguta2016}.  The same spectrum describes
the paired quarks of a two-flavour colour superconductor~\cite{Buballa2005} and
Dirac electrons in relativistic theories of
superconductivity~\cite{Ohsaku2002}.  In the nonrelativistic limit only one of
the two levels remains, and the integral becomes that of the crossover from BCS
pairing to Bose--Einstein condensation in an ordinary Fermi
gas~\cite{MariniPistolesiStrinati1998}.  We stay with pions throughout, but
what is said below about the integral applies wherever the same integral
appears.

In practice the integral is simply cut off.  As a rule the momentum
integration is stopped at $|\vec p|=\Lambda$, the cutoff is fitted together
with the coupling and the current quark mass to the vacuum, and the rest is
done numerically~\cite{HeJinZhuang2005,SunHeZhuang2007,Ayala2024}.  The cutoff
can be thought of as marking the scale where asymptotic freedom sets
in~\cite{HeJinZhuang2005}, and the physics of interest is taken to lie far
below it~\cite{SunHeZhuang2007}.  Near the onset of pion condensation it does.
At $\mu_I=4m_\pi$, however, the chemical potential of each quark, $\mu_I/2$, is
about $0.4\Lambda$ for the cutoffs of about $650$~MeV used in these works, and
the cutoff is no longer far away.  How the integral is treated near $\Lambda$
then affects the result: two treatments that describe the same vacuum lead to
visibly different pion superfluids~\cite{Avancini2019}, and
section~\ref{sec:axis} obtains both in closed form.

The difficulty is not new.  It was met early in colour superconductivity, at
the highest densities, where a sharp cutoff makes the gap vanish as the quark
chemical potential approaches the cutoff~\cite{Farias2006}.  Farias et~al.\
separated from the integral a convergent part that carries the medium, and
regularized only the rest, which depends on the vacuum
alone~\cite{Farias2006,Farias2016}.  Consistent regularizations, they argued,
should then differ only by constants~\cite{Farias2008}.  The procedure was later
named the medium separation scheme (MSS)~\cite{Farias2016}, and at isospin
density it has been used to compare the model with the lattice equation of
state that was used to build pion stars~\cite{Avancini2019}.  A second
requirement, consistency with the renormalization group, asks that the
effective action not depend on the
cutoff~\cite{BraunLeonhardtPawlowski2019}; Pannullo, Wagner and Winstel note
that in their case it is equivalent to regularizing the vacuum part
alone~\cite{PannulloWagnerWinstel2024}.  With a gap, subtracting the vacuum
limit is not enough: the remainder keeps a logarithmic divergence proportional
to the squares of the chemical potential and of the gap, and Gholami, Hofmann
and Buballa remove it by counterterms that vanish in the
vacuum~\cite{GholamiHofmannBuballa2025}.  Where different schemes have been
compared, the inhomogeneous phases and the instabilities that lead to them
depend strongly on the scheme when the whole integral is regularized, and much
less when only the vacuum part
is~\cite{PannulloWagnerWinstel2024,daSilva2026}; the comparisons of
dimensional regularization with the
cutoff~\cite{Fujihara2009,Inagaki2011,Inagaki2012} belong here as well, and
ref.~\cite{Inagaki2013} takes dimensional regularization alone to the limit of
four dimensions.  All these comparisons are numerical: the model is solved in each scheme,
and the results are set side by side.  As Pannullo et~al.\ remark, there is
still no obvious physical explanation why regularizing the vacuum part alone
should give the more trustworthy
predictions~\cite{PannulloWagnerWinstel2024}.

The starting point of this work was a calculation in one space dimension.  In
an earlier paper with Zhukovsky and
Klimenko~\cite{ZhukovskyKlimenkoKhunjua2010} we studied pion condensation in
the Gross--Neveu model (see also~\cite{EbertKlimenko2009}), where the
quasiparticles have the same two levels, and
noticed that its thermodynamic potential, the integral of their sum over a
single momentum, can be reduced to elliptic
integrals.\footnote{Equation~(18) of that paper contains a misprint.  The
results reported there were obtained numerically and do not depend on it.}
In one dimension nothing more is needed: the model is
renormalizable~\cite{GrossNeveu1974}, and with the measure $dp$ the only
divergence that depends on the state is a logarithm proportional to the square
of the total condensate, which the coupling absorbs.  In two space dimensions
the measure $p\,dp$ equals $E\,dE$, with $E$ the energy of a free quark, and the
integral is elementary.  In three the
elliptic integrals return, and with them a logarithm whose coefficient
combines the chemical potential and the condensates differently, so that the
coupling cannot absorb it.  This is the case treated here.

Elliptic integrals have appeared in pairing problems before.  The closest
precedent is the nonrelativistic limit mentioned above: in three dimensions and
at zero temperature the mean-field equations of the crossover from BCS pairing
to Bose--Einstein condensation can be solved in complete elliptic integrals of
the first and second kinds, and Marini, Pistolesi and Strinati used this to
follow analytically a crossover region that had until then been treated
numerically~\cite{MariniPistolesiStrinati1998}.  Closest in spirit is an older
work in condensed matter: Brazovskii, Dzyaloshinskii and Krichever obtained the
thermodynamics of exactly solvable Peierls chains from a differential on the
spectral curve, expanded at the point at
infinity~\cite{BrazovskiiDzyaloshinskiiKrichever1982}.  Our integral is defined
on a curve of the same kind (section~\ref{sec:cutoff}), but in a continuum and
with a uniform condensate, and the point at infinity is where its ultraviolet
divergence is located.  More distant examples are the gluon self-energy in
colour superconductors, where elliptic integrals of all three kinds
occur~\cite{Rischke2001,MalekzadehRischke2006}, and crystalline condensates in
one spatial dimension, which are described by elliptic
functions~\cite{BasarDunneThies2009}.  The tool we use is the Hermite
reduction~\cite{Hermite1883}: integration by parts that brings such an integral
down to an algebraic term and elliptic integrals of the first, second and third
kinds.  To our knowledge, however, the relativistic potential in three space
dimensions, with the mass, the gap and the chemical potential all present, has
not been written in closed form.

In this paper we evaluate the integral in closed form.  The Hermite reduction
brings it, at any cutoff and without any expansion or approximation, to an
algebraic term at the cutoff and three incomplete elliptic integrals, one of
each kind.  Their modulus and parameter, and their amplitude at the end of
the path, are given by a triangle two of whose sides are the quasiparticle
levels at zero momentum.  Integration by
parts is also the basis of differential
regularization~\cite{FreedmanJohnsonLatorre1992}, where it is formal and the
surface terms are discarded; here the boundary term is kept, and it contains
the power divergences.  The only other divergence is a logarithm, and it has a
geometric origin: the momentum axis is an open path on an elliptic curve, and
the ultraviolet is a puncture at which the path ends.

The closed form shows that a regularization is a prescription for the end of
the path.  Consider a sharp or smooth cutoff that depends only on
$|\vec p|/\Lambda$, is independent of the state, and is removed by taking
$\Lambda\to\infty$.  Its effect on the finite part is confined to three
structures, with coefficients that depend on the cutoff prescription but not on
the state (section~\ref{sec:regularization}).  One is an irrelevant additive
constant; the other two renormalize the coupling and the scale of the
logarithm.  Once the vacuum quark mass $M_0$, the pion decay constant $f_\pi$
and the pion mass $m_\pi$ fix the remaining parameters, all such momentum
cutoffs yield the same state-dependent potential, up to an additive constant.
The medium separation scheme, as implemented for this model in
ref.~\cite{Avancini2019}, reproduces this potential, up to a constant as well.

Regulators that act separately on the two levels are different.  A
proper-time kernel is one example: it generates an isospin density below the
onset of condensation and therefore violates the Silver Blaze
property~\cite{Cohen2003,Braun2021}.  Keeping a momentum cutoff finite is
different again.  At a finite cutoff the potential keeps terms of order
$\Lambda^{-2}$ that are not of these three kinds, so the result defines a
different model rather than another regularization of the same one.  This
interpretation is consistent with the proposal that, in colour
superconductivity, a low cutoff fitted to the vacuum may mimic an interaction
quenched at large momenta~\cite{Ruester2005}.  The analysis does not determine
which model is closer to QCD; it makes their difference explicit.

The paper is organized as follows.  Section~\ref{sec:model} recalls the model
and its two quasiparticle levels.  Section~\ref{sec:cutoff} reduces the
potential at any cutoff and shows where each divergence comes from.
Section~\ref{sec:regularization} asks what a regularization can change.
Section~\ref{sec:axis} follows the state along the isospin axis, obtained from
the closed form with no numerical integration over the momentum, and compares
it with the state of the model with the cutoff kept.
Section~\ref{sec:conclusions} summarizes the results.  The appendices
contain the steps of the derivation (\ref{app:identity}, \ref{app:finite}), the
comparison with the medium separation scheme (\ref{app:mss}), the derivatives
of the closed form (\ref{app:derivatives}) and the checks against known limits,
the nonrelativistic one among them (\ref{app:checks}).  The supplementary
material gives the complete derivation.

\section{The model and its two levels}\label{sec:model}
We consider the two-flavour Nambu--Jona-Lasinio model in $3+1$ dimensions, at
leading order in $1/N_c$, with a scalar and a pseudoscalar isotriplet channel
and an isospin chemical potential $\mu_I$:
\begin{equation}\label{eq:lagx}
  \mathcal L=\bar q\Bigl[\gamma^{a}i\partial_a-m_0
  +\frac{\mu_I}{2}\tau_3\gamma^{0}\Bigr]q
  +G\bigl[(\bar qq)^{2}+(\bar q\,i\gamma^{5}\vec\tau\,q)^{2}\bigr],
\end{equation}
Here $q$ is the doublet of $u$ and $d$ quarks, $\vec\tau$ are the Pauli matrices
in flavour space, $m_0$ is the current quark mass and $G$ the coupling.  Each
quark carries half of the isospin chemical potential: the $u$ and $d$ quarks
have the chemical potentials $+\nu$ and $-\nu$, with $\nu=\mu_I/2$.  In the
mean-field approximation the chiral condensate is carried by the constituent
mass $M$ and the charged pion condensate by $\Delta$; we take $M$, $\Delta$ and
$\nu$ positive, since the potential is even in $\Delta$ and in $\nu$, and write
$\rho^{2}=M^{2}+\Delta^{2}$.  The quarks then move in the single-particle
Hamiltonian
\begin{equation}\label{eq:Hmf}
  H=\gamma^{0}\boldsymbol\gamma\cdot\mathbf p+\gamma^{0}M+i\gamma^{0}\gamma^{5}\tau_1\Delta-\nu\,\tau_3 ,
\end{equation}
an $8\times8$ matrix in Dirac and flavour indices.  After the second flavour is
rotated by $i\gamma^{0}\gamma^{5}$, it takes the form of a Bogoliubov--de
Gennes Hamiltonian, whose particle block is the Dirac Hamiltonian of a quark of
mass $M$ shifted by $-\nu$ and whose pairing field is $\Delta$.  Its
characteristic polynomial is the square of a quartic $P(\eta)$, each level
appearing once for each spin projection, and the quartic factorizes:
\begin{equation}\label{eq:levels}
  P(\eta)=\bigl[\eta^{2}-(E_p+\nu)^{2}-\Delta^{2}\bigr]\bigl[\eta^{2}-(E_p-\nu)^{2}-\Delta^{2}\bigr],
  \qquad
  E^{\pm}(p)=\sqrt{(E_p\pm\nu)^{2}+\Delta^{2}} .
\end{equation}
with $E_p=\sqrt{p^{2}+M^{2}}$ and $p=|\mathbf p|$.  The eight levels are
$\pm E^{+}$ and $\pm E^{-}$, each twice, and their meaning is plain.  $E_p$ is
the energy of a free quark of mass $M$.  The pion condensate mixes the quark of
one flavour with the antiquark of the other, and the isospin potential shifts
the energies of both members of such a pair by the same amount, $-\nu$ for one
pair and $+\nu$ for the other; the condensate then enters in quadrature, as a
gap does.  $E^{-}$ is the Bogoliubov energy, with $E_p-\nu$ the energy
counted from the Fermi level and $\Delta$ the gap.

\begin{figure}[!ht]
  \centering
  \includegraphics[width=0.66\linewidth]{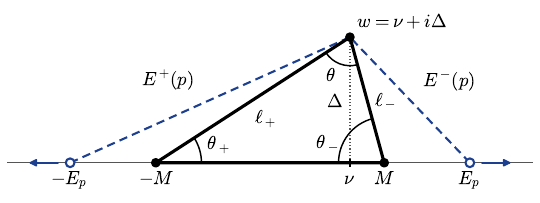}
  \caption{The two levels as two distances.  In the plane of
  $w=\nu+i\Delta$ the level $E^{\pm}(p)$ is the distance from the apex $w$ to
  the point $\mp E_p$ of the real axis (dashed).  At $p=0$ the two points are
  the ends of the base, $\mp M$, and the levels are the sides $\ell_\pm$ of the
  triangle; as $p$ grows the points slide apart.  $\theta$ is the angle at the
  apex, $\theta_\pm$ the angle between the base and the side $\ell_\pm$.}
  \label{fig:triangle}
\end{figure}

Both levels are distances.  In the plane of the complex number $w=\nu+i\Delta$,
$E^{-}$ is the distance from $w$ to the point $+E_p$ of the real axis, and
$E^{+}$ the distance from $w$ to the point $-E_p$
(figure~\ref{fig:triangle}).  At $p=0$ the two points are $\pm M$, and the
levels are the two sides of the triangle with base $[-M,M]$ and apex $w$:
\begin{equation}\label{eq:triangle}
  \ell_+\equiv E^{+}(0)=\sqrt{(M+\nu)^{2}+\Delta^{2}},\qquad
  \ell_-\equiv E^{-}(0)=\sqrt{(M-\nu)^{2}+\Delta^{2}} ,
\end{equation}
As $p$ grows the two points slide apart along the axis, and the two distances
are the levels at that momentum.  The gap of the spectrum is the distance from
the apex to the ray $[M,\infty)$: the height $\Delta$ of the triangle if its
foot $\nu$ lies on the ray, and the side $\ell_-$ if it does not.  We denote by
$\theta$ the angle of the triangle at the apex and by $\theta_\pm$ the angle
between the base and the side $\ell_\pm$.  The triangle is introduced here for
one reason: every quantity of the reduction in section~\ref{sec:cutoff} turns
out to be an element of it.

The thermodynamic potential at this order is the condensate term plus the
fermion determinant.  The sum over the Matsubara frequencies and the angular
integration are standard~\cite{Klevansky1992,Buballa2005}, and at zero
temperature, with the eight levels above, they give
\begin{equation}\label{eq:Omegax}
  \Omega=\frac{(M-m_0)^{2}+\Delta^{2}}{4G}
  -\frac{N_c}{\pi^{2}}\int_0^{\infty}\!dp\,p^{2}\,\bigl[E^{+}(p)+E^{-}(p)\bigr] .
\end{equation}
The integral diverges: at large momenta the integrand grows as $2p^{3}$.  It is
also not elementary, since $E_p$ is itself a square root and each level is a
square root of an expression containing it.  It becomes elementary when one of
$M$, $\Delta$, $\nu$ vanishes (section~\ref{sec:parts} and
appendix~\ref{app:checks}); with all three
present it is an elliptic integral, and the next section evaluates it at any
cutoff.

\section{The potential at any cutoff}\label{sec:cutoff}
\subsection{The variable}\label{sec:variable}
The integrand of \eqref{eq:Omegax} is a sum of two square roots, each with the
square root $E_p$ inside it, and that is what makes the integral hard to do in
$p$.  We integrate instead in a variable that has a polynomial relation with
$p$: the square of the integrand,
\begin{equation}\label{eq:Rdef}
  R=\bigl(E^{+}+E^{-}\bigr)^{2},
\end{equation}
the square of the sum of the two distances of figure~\ref{fig:triangle}.  The
squares of the two levels differ by $4\nu E_p$ and add up to
$2(E_p^{2}+\nu^{2}+\Delta^{2})$, so the sum obeys
$R^{2}-4(E_p^{2}+\nu^{2}+\Delta^{2})R+16\nu^{2}E_p^{2}=0$.  This contains $E_p$
only through $E_p^{2}=p^{2}+M^{2}$, and it is linear in $p^{2}$.  Its part free
of $p$ is the same equation at $p=0$, whose roots are the squares of the sum
and of the difference of the two sides of the triangle; together with the root
of the coefficient of $p^{2}$, they are the three constants of the problem,
\begin{equation}\label{eq:abc}
  a=(\ell_++\ell_-)^{2},\qquad c=(\ell_+-\ell_-)^{2},\qquad b=4\nu^{2} .
\end{equation}
In terms of them the relation reads $(R-a)(R-c)=4p^{2}(R-b)$, and solving it
for $p^{2}$ costs no radical:
\begin{equation}\label{eq:p2x}
  p^{2}=\frac{(R-a)(R-c)}{4\,(R-b)} .
\end{equation}
So $a$ and $c$ are the zeros of $p^{2}(R)$ and $b$ is its pole.  At $p=0$ the
variable starts at $R=a$, and as $p\to\infty$ it grows as $4p^{2}$: the
integration runs over $R$ from $a$ to infinity.  The three points are ordered,
$a>b>c>0$, because $(a-b)(c-b)=-16\nu^{2}\Delta^{2}$ is negative and
$ac=16M^{2}\nu^{2}$ is positive.  The medium enters through the pole alone:
its position is set by the chemical potential, and its strength by the product
$\nu^{2}\Delta^{2}$ of the potential and the condensate it pairs, as the
partial fraction~\eqref{eq:pfrac} of appendix~\ref{app:identity} shows.

With $dp=\tfrac12(p^{2})'\,dR/p$, the prime denoting $d/dR$, the measure
becomes $p^{2}\sqrt R\,dp=\tfrac12\sqrt R\,p\,(p^{2})'\,dR$, and every factor
but the product $\sqrt R\,p$ is rational in $R$.  That product is the only
irrationality left:
\begin{equation}\label{eq:psqrtR}
  \sqrt R\;p=\frac{Y}{2\,(R-b)},\qquad
  Y\equiv\sqrt{R\,(R-a)(R-b)(R-c)} ,
\end{equation}
the square root of a quartic with four distinct real roots.  The curve on which
$Y$ is single-valued therefore has genus one: it is an elliptic curve.  Three of the roots are the points \eqref{eq:abc}; the fourth, $R=0$,
comes from the integrand itself, being the point where the sum of the two
levels would vanish.  The sign of $Y$ on the path is the sign of $p$, so the
two points of the curve over $R=\infty$ are $p=+\infty$ and $p=-\infty$.  We
call them the punctures.  The momentum integral is an integral along a path on
the curve, from the branch point $R=a$, where $p=0$, to a puncture.
\subsection{Integration by parts}\label{sec:parts}
In the variable $R$ the integrand is a rational function of $R$ divided by
$Y$, and the Hermite reduction~\cite{Hermite1883} lowers it piece by piece by
integrating by parts.  Each integration by parts leaves an algebraic term at
the ends of the path.  The reason for carrying the reduction out in full is to
see what these terms are: they are the divergences.

The free case shows the idea on an answer every reader knows.  At $\nu=0$ the
two levels coincide, $E^{\pm}=\sqrt{p^{2}+\rho^{2}}$, the variable is
$R=4(p^{2}+\rho^{2})$, and the only irrationality is
$\tilde Y=\sqrt{R\,(R-4\rho^{2})}$, the square root of a quadratic.  Two
integrations by parts remove the powers $R^{2}$ and $R$ of the integrand and
leave a constant:
\begin{equation}\label{eq:free2}
  \frac{R(R-4\rho^{2})}{16}\,\frac{dR}{\tilde Y}
  =d\Bigl[\underbrace{\frac{(R-2\rho^{2})\,\tilde Y}{32}}_{W}\Bigr]
  -\frac{\rho^{4}}{8}\,\frac{dR}{\tilde Y},\qquad W\big|_{R=4\rho^{2}}=0 ,
\end{equation}
The lower limit contributes nothing, because $p=0$ is the branch point
$R=4\rho^{2}$, where $\tilde Y$ vanishes.  At the cutoff the boundary term
gives $\tfrac12\Lambda^{4}+\tfrac12\rho^{2}\Lambda^{2}+\tfrac1{16}\rho^{4}$ up to
terms that vanish as $\Lambda\to\infty$, and the remainder, through
$\int dR/\tilde Y=2\ln\bigl(\sqrt R+\sqrt{R-4\rho^{2}}\bigr)$, gives
$-\tfrac14\rho^{4}\ln(2\Lambda/\rho)$: together, the familiar vacuum energy of
a free Dirac field.  The powers of the cutoff come from the boundary term
alone, the logarithm from the remainder alone, and the lower end of the path
contributes nothing.  The general case is the same computation with a quartic
in place of the quadratic.

There the measure is $p^{2}\sqrt R\,dp=Q(R)\,dR/Y$, with $Q$ a rational
function whose only pole at finite $R$ is a double pole at $R=b$
(appendix~\ref{app:identity}).  Three integrations by parts remove its cubic
power, its square and its double pole, and what they subtract is the
derivative of one algebraic function, the boundary term
\begin{equation}\label{eq:Wdef}
  W=\Bigl[\frac1{32}+\frac{s_1}{R-b}+\frac{s_2}{(R-b)^{2}}\Bigr]\,Y ,
\end{equation}
with $s_1=(2\nu^{2}-\rho^{2})/16$ and $s_2=-\tfrac23\nu^{2}\Delta^{2}$, the index
of $s$ being the order of its pole at $b$.  What they leave has degree one and a
single simple pole:
\begin{equation}\label{eq:remx}
  Q\,\frac{dR}{Y}=dW+\Bigl[A_1+A_2\,\frac{R-c}{R-b}+A_3\,(R-b)\Bigr]\frac{dR}{Y},
\end{equation}
an identity between rational functions of $R$ once $dW$ is written out, with
the coefficients $A_1$ and $A_2$ given in appendix~\ref{app:identity}.  At the
lower limit $R=a$ the root $Y$ vanishes, so $W(a)=0$, and every algebraic term
lives at the cutoff.

The three differentials left in \eqref{eq:remx} cannot be lowered further, and
they are of the three kinds.  $dR/Y$ has no pole: it is of the first kind.
$(R-c)\,dR/[(R-b)\,Y]$ has its only pole at $R=b$; since $b$ is a branch point
of the curve, the pole is double in the local parameter there and carries no
residue, so this differential is of the second kind.  The isospin potential,
which put the pole at $b$, therefore contributes no logarithm of its own.
$(R-b)\,dR/Y$ has simple poles at the two punctures, with residues $\mp1$: it
is of the third kind, and it is a logarithm.  The only residues of the
integrand are thus at the punctures, $p=\pm\infty$, and they are $\mp A_3$, with
\begin{equation}\label{eq:residue}
  A_3=\tfrac12\nu^{2}\Delta^{2}-\tfrac18\rho^{4} ,
\end{equation}
a number attached to the differential itself, which no further manipulation
can change.  It will be met twice more: $2A_3$ is the coefficient of
$\ln\Lambda$, and also the coefficient of $p^{-3}$ in $E^{+}+E^{-}$ at large
$p$, where it is usually met.  Its two parts are the residue $-\tfrac18\rho^{4}$
of the free case~\eqref{eq:free2} and the medium,
$\tfrac12\nu^{2}\Delta^{2}$, half the strength of the pole of $p^{2}(R)$.
\subsection{Legendre's form, the triangle and the torus}\label{sec:torus}
The three integrals left are brought to Legendre's normal form by the
substitution of Byrd and Friedman~\cite{ByrdFriedman1971} (appendix~\ref{app:identity},
\eqref{eq:legsub}): with $\sn u=\sin\varphi$ they become the incomplete
integrals $\EllF$, $\EllE$ and $\Pi$ of modulus $k$, with the parameter
$\alpha^{2}$ in $\Pi$ and a common factor $g$~\eqref{eq:legdict}.  The upper
end of the path, $R\to\infty$, is at $\sin^{2}\varphi_1=1/\alpha^{2}$, exactly
where the integral of the third kind has its pole: the puncture, seen from the
normal form.  The Legendre data are elements of the triangle of
figure~\ref{fig:triangle}:
\begin{equation}\label{eq:tridict}
  k=\cos\frac{\theta}{2},\qquad
  \sin\varphi_1=\frac{2\nu}{\ell_++\ell_-},\qquad
  g=\frac{1}{2\nu\sqrt{\ell_+\ell_-}},\qquad
  \omega^{2}\equiv\frac{k^{2}}{\alpha^{2}}=\frac{4\nu^{2}-(\ell_+-\ell_-)^{2}}{4\,\ell_+\ell_-} ,
\end{equation}
the last being the parameter that appears in the finite part.  The modulus is
the cosine of half the angle at the apex; in the angles at the base,
$k=\sin\tfrac12(\theta_-+\theta_+)$ and $\omega=\sin\tfrac12(\theta_--\theta_+)$.
Since $\ell_+^{2}-\ell_-^{2}=4M\nu$, the sine of the amplitude of the endpoint
is also $(\ell_+-\ell_-)/(2M)$, which in the angles of the triangle is
$\omega/k$.  The modulus, the parameter and the amplitude of the endpoint are
thus fixed by the angles of the triangle: they give its shape, and $g$, the
only one of the four with a dimension, gives its size, to the power $-2$.  The
ends of the range of the modulus are reached when the triangle collapses.  As
$\Delta\to0$ at $\nu<M$ the apex descends onto the base, $\theta\to\pi$ and
$k\to0$.  The angle at the apex closes, $\theta\to0$ and $k\to1$, in two ways:
as $M\to0$ the base shrinks to a point, and as $\Delta\to0$ at $\nu>M$ the apex
descends onto the line of the base beyond its end.  All three limits are
taken in appendix~\ref{app:checks}.

\begin{figure}[!ht]
  \centering
  \includegraphics[width=\linewidth]{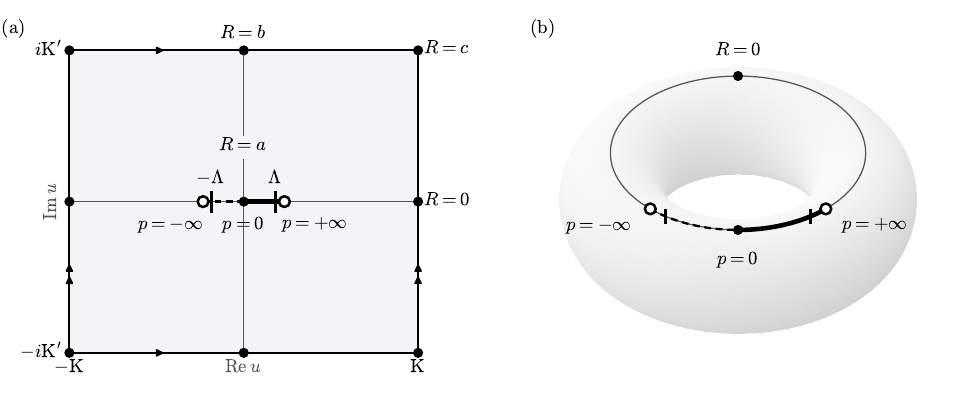}
  \caption{The curve $Y^{2}=P_4(R)$ as a torus, for the triangle of
  Fig.~\ref{fig:triangle}.  (a) The rectangle of periods of the coordinate
  $u$; opposite sides are identified.  The four half-periods are the four
  branch points $R=a,0,c,b$.  The open circles are the two punctures,
  $R=\infty$; the segment between them is the momentum line, the half along
  which the integral runs drawn thick; the bars show where a cutoff
  $\Lambda=2M$ stops it.  (b) The same on the torus: the real axis of (a) is
  the circle on top, and the whole momentum line is a short arc of it.  The
  points $R=b$ and $R=c$ are on the underside.}
  \label{fig:torus}
\end{figure}

The substitution has a geometric meaning.  Since
$g\,u=\int_a^{R}dR/Y$, the variable $u$ is the integral of the differential of
the first kind, which is finite everywhere on the curve, and so it is a
coordinate on the whole curve.  $R$ is a rational function of $\sn^{2}u$; it
is even in $u$ and has the two periods $2\EllK$ and $2i\EllK'$, with
$\EllK=\EllK(k)$ and $\EllK'=\EllK(k')$ the complete integrals of the first
kind.  A plane in which two periods are identified is a torus: the curve is the
rectangle of figure~\ref{fig:torus}a with its opposite sides glued together,
figure~\ref{fig:torus}b.  The points $u$ and $-u$ have the same $R$ and
opposite $Y$; they coincide at the four half-periods, and these are the four
branch points,
\begin{equation}\label{eq:halfperiods}
  u=0,\quad \EllK,\quad \EllK+i\EllK',\quad i\EllK'\qquad\text{are}\qquad R=a,\quad 0,\quad c,\quad b .
\end{equation}
The two punctures are the points $u=\pm u_1$ of the real axis, with
$\sn u_1=\sin\varphi_1$.  Since the sign of $Y$ on the path is the sign of $p$,
the segment $-u_1<u<u_1$ is the whole momentum line, $-\infty<p<\infty$, with
$p=0$ at its centre, and the integral~\eqref{eq:Omegax} runs along half of it.

The picture says three things.  The momentum line is open: it joins the two
punctures and does not go round the torus, which is why the result is made of
incomplete integrals, taken to the amplitude $\varphi_1$, and not of periods.
The ultraviolet is local: a cutoff on the momentum stops the path short of the
puncture, and to remove the cutoff is to go all the way to it, so that whatever
diverges is decided by the behaviour of the integrand at the puncture and
nowhere else.  And the triangle fixes the torus: the shape of the rectangle,
$\EllK'/\EllK$, depends only on the angle $\theta$ at the apex, and the
punctures sit at the amplitude $\varphi_1$.  A spectral curve whose point at
infinity carries the thermodynamics is met also in the exactly solvable
Peierls models of Brazovskii, Dzyaloshinskii and
Krichever~\cite{BrazovskiiDzyaloshinskiiKrichever1982}.

On the torus the three survivors of \eqref{eq:remx} are the three kinds of
differential it carries: $du$ itself, with no pole; one with a double pole at
the half-period $i\EllK'$, which is $R=b$; and one with simple poles at the two
punctures.  The first two are finite along the whole path.  The third is not,
and what it does at the end of the path is the subject of the next subsection.
\subsection{The identity at any cutoff}\label{sec:identity}
Collecting the boundary term, the three remaining integrals and their Legendre
forms, we obtain
\begin{equation}\label{eq:identity}
  \begin{split}
  \int_0^{\Lambda}p^{2}\,\bigl[E^{+}+E^{-}\bigr]\,dp
  &=W(R_\Lambda)+g\Bigl[A_1\,\EllF(\varphi_\Lambda,k)
    +A_2\,\frac{a-c}{a-b}\,\EllE(\varphi_\Lambda,k)\\
  &\hphantom{{}=W(R_\Lambda)+g\Bigl[}
    +A_3\,(a-b)\,\Pi(\varphi_\Lambda;\alpha^{2},k)\Bigr],\\
  \sin^{2}\varphi_\Lambda&=\frac{b\,(R_\Lambda-a)}{a\,(R_\Lambda-b)},
  \end{split}
\end{equation}
exact at every $\Lambda$, with nothing expanded or dropped.  The left side is
the momentum integral of \eqref{eq:Omegax} cut off at $\Lambda$; the right side
is made of three algebraic coefficients, three elliptic integrals and the three
points.  Every symbol on the right is an explicit function of $M$, $\Delta$,
$\nu$ and $\Lambda$: the points by \eqref{eq:abc}, the end of the path
$R_\Lambda=[E^{+}(\Lambda)+E^{-}(\Lambda)]^{2}$ and the amplitude
$\varphi_\Lambda$ by the second line, the boundary term by \eqref{eq:Wdef}, the
coefficients by \eqref{eq:residue} and appendix~\ref{app:identity}.  No
equation has to be solved to evaluate it.  Inserted into \eqref{eq:Omegax},
\eqref{eq:identity} is the thermodynamic potential of the model with a sharp
cutoff on the three-momentum, in closed form.

At large $\Lambda$ each term can be expanded.  The algebraic terms are Laurent
series in $1/\sqrt{R_\Lambda}$, re-expanded in $\Lambda$.  The integrals of the
first and second kinds tend to their values at the amplitude $\varphi_1$ of
the puncture, with corrections of order $\Lambda^{-2}$.  The integral of the
third kind diverges logarithmically, because its pole sits at the end of the
path (appendix~\ref{app:finite}).  Term by term:
\begin{equation}\label{eq:whence}
  {\renewcommand{\arraystretch}{1.45}
  \begin{array}{l|ccc|c}
  & \Lambda^{4} & \Lambda^{2} & \Lambda^{0} & \ln\Lambda\\ \hline
  \tfrac1{32}\,Y_\Lambda & \tfrac12 & \tfrac34\rho^{2}-\tfrac12\nu^{2}
     & \tfrac3{16}\rho^{4}+\tfrac14\nu^{2}(2\Delta^{2}-M^{2}) & \\
  s_1\,Y_\Lambda/(R_\Lambda-b) & & \tfrac12\nu^{2}-\tfrac14\rho^{2}
     & \tfrac18\rho^{2}(2\nu^{2}-\rho^{2}) & \\
  s_2\,Y_\Lambda/(R_\Lambda-b)^{2} & & & -\tfrac23\nu^{2}\Delta^{2} & \\
  g\,A_1\,\EllF & & & g\,A_1\,\EllF(\varphi_1,k) & \\
  g\,A_2\,\tfrac{a-c}{a-b}\,\EllE & & & g\,A_2\,\tfrac{a-c}{a-b}\,\EllE(\varphi_1,k) & \\
  g\,A_3\,(a-b)\,\Pi & & & \text{finite part: next section} & 2A_3\\ \hline
  \text{sum} & \tfrac12 & \tfrac12\rho^{2} & \mathcal C_W+\text{elliptic} & 2A_3
  \end{array}}
\end{equation}
Four things can be read off the table.  The quartic divergence comes from the
bare term alone, with the coefficient of the free case.  The quadratic one is
$\tfrac12\rho^{2}\Lambda^{2}$ only in the sum: the bare term carries
$\tfrac34\rho^{2}-\tfrac12\nu^{2}$, the term with the simple pole returns
$\tfrac12\nu^{2}-\tfrac14\rho^{2}$, and the chemical potential cancels, so that
no power of the cutoff comes with a dependence on $\nu$.  The constant
$\mathcal C_W$ of the boundary term is the sum of the three algebraic entries
of the third column.  And the logarithm comes from the third kind alone, with
twice the residue \eqref{eq:residue} as its coefficient: it is the only
divergence that depends on the medium.  In sum,
\begin{equation}\label{eq:asymp}
  \int_0^{\Lambda}p^{2}\,\bigl[E^{+}+E^{-}\bigr]\,dp
  =\frac{\Lambda^{4}}{2}+\frac{\rho^{2}\Lambda^{2}}{2}+2A_3\ln\Lambda
  +\mathcal C+O(\Lambda^{-2}),
\end{equation}
with a finite part $\mathcal C$ made of $\mathcal C_W$, of
$\EllF(\varphi_1,k)$ and $\EllE(\varphi_1,k)$, and of what remains of the
integral of the third kind at its pole.  The last piece is the only place where
a relation from the tables is needed, and section~\ref{sec:finitepart} writes
$\mathcal C$ in closed form.
\subsection{The finite part}\label{sec:finitepart}
What remains is the finite part of the integral of the third kind at its
pole.  For it one relation from the tables is needed, the change of parameter
of the third kind~\cite{DLMF}.  It exchanges the pole at $\alpha^{2}$ for a pole
at $\omega^{2}=k^{2}/\alpha^{2}$, away from the end of the path, at the price of
an elementary function; at the end of the path the elementary function
degenerates, and it carries the logarithm (appendix~\ref{app:finite}).  With
it, and with the first two integrals at the amplitude $\varphi_1$ of the
puncture, the finite part of \eqref{eq:asymp} is
\begin{equation}\label{eq:Cfin}
  \begin{split}
  \mathcal C&=\mathcal C_W
  +g\,A_1\,\EllF(\varphi_1,k)+g\,A_2\,\frac{a-c}{a-b}\,\EllE(\varphi_1,k)\\
  &\quad+g\,(a-b)\,A_3\,\bigl[\EllF(\varphi_1,k)-\Pi(\varphi_1;\omega^{2},k)\bigr]
  +A_3\ln\frac{16}{a+b-c},\\
  \sin^{2}\varphi_1&=\frac{b}{a},\qquad
  k^{2}=\frac{a\,(b-c)}{b\,(a-c)},\qquad
  \omega^{2}=\frac{b-c}{a-c},\qquad
  g=\frac{2}{\sqrt{b\,(a-c)}} ,
  \end{split}
\end{equation}
where the new integral of the third kind is finite, since
$\omega^{2}\sin^{2}\varphi_1=b(b-c)/[a(a-c)]<1$.  Three functions ---
$\EllF$, $\EllE$ and one $\Pi$, all incomplete Legendre integrals at the same
amplitude $\varphi_1$ --- and one logarithm carry the whole transcendental
content of the potential.  Neither logarithm has a dimensionless argument by
itself, but together they have, as they must: since
$16/(a+b-c)=4/(\ell_+\ell_-+\nu^{2})$, the logarithms of \eqref{eq:asymp} and
\eqref{eq:Cfin} combine into
$2A_3\ln\bigl(\Lambda\big/\tfrac12\sqrt{\ell_+\ell_-+\nu^{2}}\bigr)$, whose
scale is set by the triangle itself and is $\rho/2$ at $\nu=0$, as in the free
case.  Everything in \eqref{eq:Cfin} is a known function of the three points,
and the three points are known functions of $M$, $\Delta$ and $\nu$
by~\eqref{eq:abc}: this is the closed form of the finite part.  Which of its
terms a regularization can change is the subject of the next section.

\section{What a regularization can change}\label{sec:regularization}
\subsection{Rules that know only the momentum}\label{sec:momentum}
The reduction has put everything that diverges at one end of the path: in the
boundary term taken there and in one logarithm.  A regularization is
therefore a rule for that end, which says where the integral is stopped or how
the integrand is made to fade as the momentum grows.  Stopped sharply at
$\Lambda$, the integral is \eqref{eq:asymp}: three functions of one scale carry
everything that grows, and the finite part $\mathcal C$ is the only term that
does not depend on where the path is stopped.  Another rule puts its own scale
in the place of $\Lambda$, and what it does to the finite part is decided by
the three coefficients $\tfrac12$, $\tfrac12\rho^{2}$ and $2A_3$.

What may the rule know?  If it is to belong to the model and to be the same
for every state, it may know the momentum and nothing about the state.  The
three coefficients it meets need none of the reduction.  They are the
coefficients of the sum of the two levels expanded at large momenta, in odd
powers of $p$:
\begin{equation}\label{eq:puncture}
  E^{+}+E^{-}=2p+\frac{\rho^{2}}{p}+\frac{\nu^{2}\Delta^{2}-\rho^{4}/4}{p^{3}}+O(p^{-5}) .
\end{equation}
Multiplied by $p^{2}$ and integrated to $\Lambda$, its three terms give the
three divergences of \eqref{eq:asymp}, and the rest of the integrand is
integrable to infinity.  The last coefficient is the $2A_3$ of
section~\ref{sec:parts}: what the reduction found at the puncture is the
coefficient of the local expansion there.  A rule that acts on the momentum
alone sees these three coefficients and nothing else.

Two ways of changing such a rule show how much it can do.  A rescaling of the
cutoff by a constant factor multiplies the two powers by constants and adds
$2A_3$ times the logarithm of the factor to the finite part.  A smooth cutoff
function $\chi(p/\Lambda)$ in place of the sharp one, a form factor with
$\chi(0)=1$ and fast decay, changes the integral by an amount that comes from
momenta of order $\Lambda$, where \eqref{eq:puncture} holds.  In the limit only
the three terms of the expansion contribute, each through one moment of the
function in the variable $x=p/\Lambda$.  For the sharp cutoff the first two
moments are $\int_0^{1}2x^{3}\,dx=\tfrac12$ and $\int_0^{1}x\,dx=\tfrac12$, the
coefficients of the two powers.  For the Gaussian, $\chi=e^{-x^{2}}$, the
three moments are
\begin{equation}\label{eq:window}
  \int_0^{\infty}2x^{3}\chi\,dx=1,\qquad
  \int_0^{\infty}x\chi\,dx=\frac12,\qquad
  \int_0^{\infty}\bigl[\chi-\theta(1-x)\bigr]\frac{dx}{x}=-\frac{\gamma_E}{2} ,
\end{equation}
the third taken relative to the sharp cutoff.  The Gaussian thus doubles the
$\Lambda^{4}$ term, leaves the $\Lambda^{2}$ term as it was, and shifts the
finite part by $-\gamma_E A_3$.  The powers leave a trace of their own when
the cutoff is redefined by an amount of order $1/\Lambda$: finite terms
proportional to $1$ and to $\rho^{2}$.  In every case the finite part moves
within $\operatorname{span}\{1,\rho^{2},A_3\}$, with coefficients that depend
on the rule and not on the state.

Not every term of the finite part can be moved in this way.  The constant of
the boundary term, the sum of the algebraic entries of the column
$\Lambda^{0}$ in \eqref{eq:whence}, is not in the span:
\begin{equation}\label{eq:polspan}
  \mathcal C_W\notin\operatorname{span}\{1,\rho^{2},A_3\},\qquad
  \mathcal C_W=-\frac{A_3}{2}+\frac{\nu^{2}\Delta^{2}}{3} ,
\end{equation}
because $\nu^{2}\Delta^{2}$, as a function of the state, is not a combination
of the three.  The first part of $\mathcal C_W$ is a convention: it is what a
rescaling of the cutoff by $e^{-1/4}$ adds.  The second is not.  A rule that
knows only the momentum can move the finite part along $A_3$, in which
$\nu^{2}\Delta^{2}$ and $\rho^{4}$ come in a fixed proportion; once one vacuum
condition fixes the scale of the logarithm, both parts of $A_3$ are fixed
together.  The proportion has a familiar origin.  When the model is rewritten in
terms of mesons, their kinetic term and their quartic coupling come with one
and the same logarithmic integral $I_0$~\cite{Eguchi1976}, which is how the
compositeness condition appears there~\cite{Shizuya1980}.  The isospin
potential enters the time derivative of the charged pions, so the kinetic term
gives $\tfrac12\mu_I^{2}\Delta^{2}$ with the same $I_0$, and the logarithmic part
of the potential is $I_0(\tfrac12\mu_I^{2}\Delta^{2}-\tfrac12\rho^{4})=4I_0A_3$.  The finite part
modulo the span is therefore the same for all these rules, and it is this,
taken as a whole, that is a prediction and not a convention.  The constant
$\mathcal C_W$ contributes to it on the same footing as the three elliptic
terms of \eqref{eq:Cfin}: its part outside the span, $\tfrac13\nu^{2}\Delta^{2}$ in
the basis of \eqref{eq:remx}, cannot be discarded with the conventions.
A counterterm that knows the medium is not a rule of this kind.  Gholami,
Hofmann and Buballa remove the logarithm proportional to the squares of the
chemical potential and of the gap by counterterms that vanish in the
vacuum~\cite{GholamiHofmannBuballa2025}; such a term moves the finite part
along $\nu^{2}\Delta^{2}$ by itself, and the coefficient of $\nu^{2}\Delta^{2}$
then becomes a convention as well.
\subsection{Rules that act on the levels}\label{sec:levels}
A rule that acts on the energies, on each of the two levels by itself, is of
another kind.  At large momenta the levels are
\begin{equation}\label{eq:energyreg}
  E^{\pm}=p\pm\nu+\frac{\rho^{2}}{2p}+O(p^{-2}) :
\end{equation}
each level remembers the chemical potential at order $p^{0}$, while their sum
\eqref{eq:puncture} has forgotten it.  A cutoff function on the energies,
$E^{+}\chi(E^{+}/\Lambda)+E^{-}\chi(E^{-}/\Lambda)$, therefore weighs each
level by its own shift.  With $q=p\pm\nu$, so that $p^{2}=(q\mp\nu)^{2}$, the
integral
$\int p^{2}(p\pm\nu)\,\chi\bigl((p\pm\nu)/\Lambda\bigr)\,dp$ contains
$\nu^{2}\!\int q\,\chi(q/\Lambda)\,dq=\nu^{2}\Lambda^{2}\!\int_0^{\infty}x\chi\,dx$,
the second moment of \eqref{eq:window} again, and the two levels add up to
$2\nu^{2}\Lambda^{2}\!\int_0^{\infty}x\chi\,dx$.  For the proper-time kernel,
$\chi(x)=-\Gamma(-\tfrac12,x^{2})/(2\sqrt\pi)$, the moment is $-\tfrac14$ and
the term is $-\tfrac12\nu^{2}\Lambda^{2}$: a quadratic divergence that depends
on the chemical potential, and whose derivative with respect to $\nu$ is a
density.

It is a density the regulator invented.  In the vacuum, where $\Delta=0$ and
$\nu<M$, the two levels are $E_p\pm\nu$, their sum does not contain $\nu$, and
at zero temperature nothing depends on the isospin chemical potential below
the onset of condensation.  This is the Silver Blaze
property~\cite{Cohen2003}, and a regulator that acts on the energies violates
it.  In the functional renormalization group the same property has been
stated as a requirement on the regularization scheme, and regulators built
around the Fermi surface have been found to violate it~\cite{Braun2021}.  No
rule that knows only the momentum produces such a term: in the column
$\Lambda^{2}$ of \eqref{eq:whence} the chemical potential cancels.
\subsection{The potential with the cutoff removed}\label{sec:removed}
The coefficients that a rule puts at $1$, $\rho^{2}$ and $A_3$ go to three
different places.  That of $1$
only adds a constant to the potential; that of $\rho^{2}$ goes into the
coupling, and that of $A_3$ into the scale of the logarithm.  The potential can
therefore be written once for all the rules of
section~\ref{sec:momentum}, and its simplest form is the one with the cutoff
removed to infinity.

At large momenta half the sum of the two levels is, to the order that matters,
the energy of a free particle of mass $\rho$: by \eqref{eq:puncture},
$\tfrac12(E^{+}+E^{-})=E_\rho+\kappa/(2p^{3})+O(p^{-5})$, with
$E_\rho=\sqrt{p^{2}+\rho^{2}}$ and $\kappa=\nu^{2}\Delta^{2}$.  Two elementary
terms therefore carry the whole asymptotics of the integrand.  The first,
$2p^{2}E_\rho=2p^{3}+\rho^{2}p-\tfrac14\rho^{4}/p+\dots$, is the free case of
section~\ref{sec:parts} at the mass $\rho$; the second,
$\kappa\,p^{2}/E_\rho^{3}=\kappa/p+\dots$, is the medium; together they give
$2p^{3}+\rho^{2}p+2A_3/p$.  Their integrals to $\Lambda$ are elementary (the
second is \eqref{eq:carriers}), and the difference between the integrand and
these two carriers is integrable to infinity.  Dropping the $\Lambda^{4}$ and
the constants, absorbing the coefficient of $\rho^{2}\Lambda^{2}$ into the
coupling, and replacing the cutoff in the logarithm by a scale $\lambda$, we
obtain the potential with the cutoff removed,
\begin{equation}\label{eq:Omclass}
  \Omega_\infty=\frac{A'}{2}\,\rho^{2}-B'M-\frac{N_c}{\pi^{2}}\,\mathcal F,\qquad
  A'=\frac1{2G}-\frac{N_c\Lambda^{2}}{\pi^{2}},\quad B'=\frac{m_0}{2G},
\end{equation}
where $A'$ is written for the sharp cutoff and absorbs, for any other rule,
that rule's coefficient of $\rho^{2}\Lambda^{2}$.  The cutoff is thus removed
from everything but the logarithm, where it survives as the scale $\lambda$.
The three numbers of the model, fixed by the vacuum below, are $A'$, $B'$ and
$\lambda$, and the function of the state is
\begin{equation}\label{eq:Fclass}
  \begin{split}
  \mathcal F&=-\frac{\rho^{4}}{4}\ln\frac{2\lambda}{\rho}+\frac{\rho^{4}}{16}
  +\kappa\Bigl(\ln\frac{2\lambda}{\rho}-1\Bigr)+\mathcal R
  =\mathcal C+2A_3\ln\lambda ,\\
  \mathcal R&=\int_0^{\infty}\Bigl[p^{2}\bigl(E^{+}+E^{-}\bigr)-2p^{2}E_\rho-\frac{\kappa\,p^{2}}{E_\rho^{3}}\Bigr]dp .
  \end{split}
\end{equation}
The second equality is \eqref{eq:asymp} with the carriers subtracted on both
sides: their logarithms, $-\tfrac14\rho^{4}\ln\Lambda$ and $\kappa\ln\Lambda$,
add up to $2A_3\ln\Lambda$.  The first equality is how the potential is evaluated when
the elliptic functions are not wanted: the two carriers by hand, the remainder
$\mathcal R$ by a convergent quadrature.  The closed form \eqref{eq:Cfin}
gives the same number.  In the language of section~\ref{sec:torus}, the
subtraction of the carriers removes the pole at the puncture, and
\eqref{eq:Cfin} says in closed form what it leaves.

The three numbers are fixed in the vacuum, where $\nu=\Delta=0$ and the
triangle is its base alone.  The integrand is then $2p^{2}E_p$, so that
$\mathcal R=0$ and
$\mathcal F=-\tfrac14M^{4}\ln(2\lambda/M)+\tfrac1{16}M^{4}$, and the vacuum
mass $M_0$ obeys
\begin{equation}\label{eq:gapvac}
  A'M_0-B'=\frac{N_c}{\pi^{2}}\,M_0^{3}\Bigl[\frac12-\ln\frac{2\lambda}{M_0}\Bigr] .
\end{equation}
The pion decay constant, taken at zero momentum of the pion, is the integral
of the second carrier at $\rho=M_0$,
$f_\pi^{2}=(N_cM_0^{2}/2\pi^{2})\int_0^{\Lambda}p^{2}dp/E_p^{3}$; with
$\lambda$ in place of the cutoff, as in \eqref{eq:Fclass},
\begin{equation}\label{eq:fpivac}
  f_\pi^{2}=\frac{N_cM_0^{2}}{2\pi^{2}}\Bigl[\ln\frac{2\lambda}{M_0}-1\Bigr] ,
\end{equation}
which fixes $\lambda$ from $f_\pi$ and $M_0$.  The pion mass enters through the
onset of pion condensation, which the mean field places at $\mu_I=m_\pi$
exactly: the coefficient of $\Delta^{2}$ in $\Omega$ vanishes where the pion
propagator of the same approximation has its pole, and the two conditions are
one integral~\cite{HeJinZhuang2005}.  In the integrand that coefficient is
$p^{2}E_p/(E_p^{2}-\nu^{2})$, which is again the two carriers, now at
$\Delta=0$, and a remainder that converges and is elementary.  With the
carriers at the scale $\lambda$, the condition at $\nu=m_\pi/2$ gives $A'$, and
\eqref{eq:gapvac} then gives $B'$:
\begin{equation}\label{eq:thouless}
  \begin{split}
  A'&=\frac{N_c}{2\pi^{2}}\Bigl[M_0^{2}-\bigl(2M_0^{2}-m_\pi^{2}\bigr)\ln\frac{2\lambda}{M_0}
  -m_\pi\sqrt{4M_0^{2}-m_\pi^{2}}\,\arcsin\frac{m_\pi}{2M_0}\Bigr],\\
  B'&=\frac{N_c}{2\pi^{2}}\,M_0\,m_\pi\Bigl[m_\pi\ln\frac{2\lambda}{M_0}
  -\sqrt{4M_0^{2}-m_\pi^{2}}\,\arcsin\frac{m_\pi}{2M_0}\Bigr] .
  \end{split}
\end{equation}
In the chiral limit $m_\pi=0$, and $B'$ vanishes with it.  Three conditions,
\eqref{eq:gapvac}, \eqref{eq:fpivac} and \eqref{eq:thouless}, fix the three
numbers $A'$, $B'$, $\lambda$ from $M_0$, $f_\pi$ and $m_\pi$, each in closed
form.  After them nothing is left free, and every number of
section~\ref{sec:axis} is a prediction.

The model as it is traditionally regularized, with the sharp cutoff on the
whole integral kept at a finite $\Lambda$, uses the same three inputs to fix
$G$, $m_0$ and $\Lambda$.  At the same three numbers its potential differs
from \eqref{eq:Omclass} by the tail of order $\Lambda^{-2}$ in
\eqref{eq:asymp}, which is known explicitly, term by term
(appendix~\ref{app:finite}).  Fitted to the same vacuum, the tail moves the
three numbers as well: in the logarithm the model with the cutoff kept has its
$\Lambda=659.4$~MeV where \eqref{eq:Omclass} has $\lambda=759$~MeV.  The two
are different models, not two regularizations of one, and
section~\ref{sec:axis} shows the size of the difference on the condensates.
\subsection{The medium separation scheme gives the same potential}\label{sec:msssub}
Subtracting carriers of this kind and integrating the remainder to infinity
is what the medium separation scheme does.  In the form in which it has been
applied to this model, equation~(2.28) of ref.~\cite{Avancini2019}, the
carriers are expanded about a fixed mass $M_s$, in the reference the vacuum
mass, instead of being taken at $\rho$.  The cutoff then enters through two
vacuum integrals, $I_{\rm q}=\int_0^{\Lambda}p^{2}dp/E_s$ and
$I_{\ln}=\int_0^{\Lambda}p^{2}dp/E_s^{3}$ with $E_s=\sqrt{p^{2}+M_s^{2}}$, and
nowhere else; the remainder converges and is integrated over all momenta
(appendix~\ref{app:mss}).  Its value follows from \eqref{eq:asymp} as that of
$\mathcal R$ did, and the potential of the scheme becomes
\begin{equation}\label{eq:mssmap}
  \begin{split}
  &\Omega_{\rm MSS}=\Omega_\infty+\Omega_s\qquad\text{with}\\
  &\ln\frac{2\lambda}{M_s}=1+I_{\ln},\qquad
  A'=\frac1{2G}-\frac{2N_c}{\pi^{2}}\Bigl[I_{\rm q}+\frac{M_s^{2}}{2}\,I_{\ln}+\frac{M_s^{2}}{4}\Bigr],\qquad
  B'=\frac{m_0}{2G},\\
  &\Omega_s=\frac{m_0^{2}}{4G}+\frac{N_c}{\pi^{2}}\Bigl[M_s^{2}I_{\rm q}+\frac{M_s^{4}}{4}\,I_{\ln}+\frac{M_s^{4}}{16}\Bigr] .
  \end{split}
\end{equation}
The identity is exact and holds at every state: the potential of the scheme is
the potential \eqref{eq:Omclass} with these three numbers, plus a constant
$\Omega_s$ that no derivative sees.  It does not depend on how the two vacuum
integrals are evaluated.  Whatever values they are given, by a cutoff or in
another way, they enter only as $A'$ and $\lambda$.  With the sharp cutoff and
$\Lambda\gg M_s$, $\lambda=\Lambda\,[1+O(M_s^{2}/\Lambda^{2})]$ and $A'$ is the
$A'$ of \eqref{eq:Omclass}: the scheme is then the sharp cutoff sent to
infinity.

The scheme is therefore equivalent to \eqref{eq:Omclass}: it fixes the three
numbers through the parameters $G$, $m_0$ and $\Lambda$ of the
model with the cutoff kept, where section~\ref{sec:removed} took them from
$M_0$, $f_\pi$ and $m_\pi$ directly.  With the subtraction made at the vacuum
mass, the scheme has the vacuum mass and the pion decay constant of the cutoff
model with the same parameters.  Its onset of pion condensation comes slightly
earlier, because one integral of the onset condition is no longer cut: by
$0.4$~MeV in $\mu_I$ for the model of section~\ref{sec:axis}
(appendix~\ref{app:mss}).

\section{The state along the isospin axis}\label{sec:axis}
At given $\nu$ the state is the local minimum of \eqref{eq:Omclass} that
continues the vacuum:
\begin{equation}\label{eq:thermo}
  \partial_M\Omega=\partial_\Delta\Omega=0,\qquad
  n_I=-\frac{\partial\Omega}{\partial\mu_I}=-\tfrac12\,\partial_\nu\Omega,
\end{equation}
first derivatives all, and by the gap equations the total derivative of
$\Omega$ along the equilibrium is the partial one.  In
figure~\ref{fig:triangle}, $\nu$ is where the apex stands over the base, and
the two gap equations choose the rest of the triangle: its half-base $M$ and
its height $\Delta$.  Differentiation does not lead out of the closed form.
The derivatives of $\EllF$, $\EllE$ and $\Pi$ with respect to their amplitude,
modulus and parameter close on the same three functions, so every derivative
of the potential, of any order, is again made of the three Legendre integrals,
the logarithm and algebraic terms (appendix~\ref{app:derivatives}).  The
minimum is a local one.  The potential \eqref{eq:Omclass} is not bounded
below, because the term $(N_c/4\pi^{2})\,\rho^{4}\ln(2\lambda/\rho)$ that
\eqref{eq:Fclass} puts into it falls as $-\rho^{4}\ln\rho$ at large $\rho$ and
outgrows the rest.

\begin{figure}[!t]
  \centering
  \includegraphics[width=0.96\linewidth]{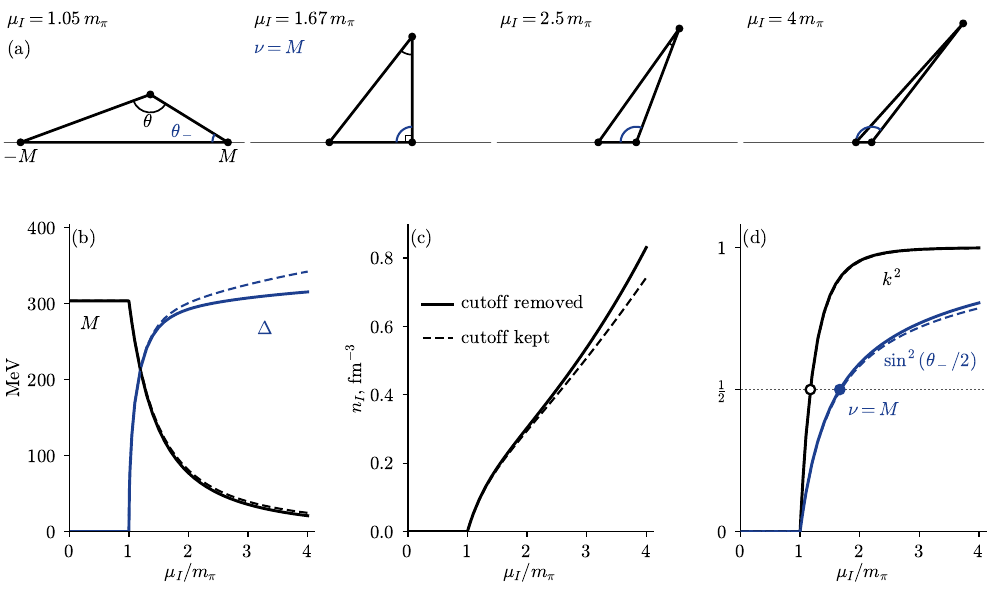}
  \caption{The state along the isospin axis with the cutoff removed, for the numbers of
  \eqref{eq:classnum}.  (a) The triangle at four values of $\mu_I=2\nu$, all on
  one scale, with the angle $\theta$ at its apex and the angle $\theta_-$ at
  the end $+M$ of its base.  At the onset the triangle is flat; then the apex
  rises and the base shrinks.  In the second frame $\nu=M$: the apex stands
  above the end of the base, and the angle there is a right angle.  (b) The
  constituent mass $M$ and the condensate $\Delta$.  (c) The isospin density.
  (d) The two angles along the axis: $k^{2}=\cos^{2}(\theta/2)$ for the angle
  at the apex, and $\sin^{2}(\theta_-/2)$ for the angle at the base.  The
  open circle marks the right angle at the apex, the full circle the right
  angle at the base.  Solid lines: the cutoff removed.  Dashed lines: the sharp
  cutoff kept at $\Lambda=659.4$~MeV, with the same vacuum.}
  \label{fig:state}
\end{figure}

We compare two models that differ only in what is done at the end of the
path.  The first is the potential with the cutoff removed.  The three vacuum
conditions \eqref{eq:gapvac}--\eqref{eq:thouless}, with $M_0=303.5$,
$f_\pi=92.4$ and $m_\pi=131.7$~MeV, give
\begin{equation}\label{eq:classnum}
  \lambda=759.076~\text{MeV},\qquad A'=-0.029425~\text{GeV}^{2},\qquad B'=0.00050073~\text{GeV}^{3},
\end{equation}
and with them \eqref{eq:Omclass} is a closed-form function of $M$, $\Delta$
and $\nu$: three Legendre integrals, one logarithm and algebra.  On the axis
$\nu=\Delta=0$ its vacuum is separated from the fall at large $\rho$ by a
barrier at $M=784$~MeV, far above the states of figure~\ref{fig:state}.  The
second is the model
with the sharp cutoff kept, at $\Lambda=659.4$~MeV, $G=4.780~\text{GeV}^{-2}$
and $m_0=4.74$~MeV, fitted to the same vacuum (its vacuum mass is
$303.6$~MeV, and its onset is at $\mu_I=131.5$~MeV).  By \eqref{eq:identity}
it is a closed form as well, with the integrals taken at $\varphi_\Lambda$
instead of $\varphi_1$.  For each model the two gap equations
\eqref{eq:thermo} are solved at each $\nu$, and no momentum integral is
computed at any step.  The derivatives are taken by differentiating the closed
forms numerically, at twenty digits; the rules of
appendix~\ref{app:derivatives} give the same derivatives in closed form.
Up to the fit of the vacuum, the solid curves are the medium separation scheme
and the dashed ones the traditional cutoff at the parameters of
ref.~\cite{Avancini2019}, whose figure~1 shows the same two curves for
$\Delta$ (its horizontal axis is our $\mu_I/m_\pi$); ours agree with it.

Figure~\ref{fig:state} shows the result.  Pion condensation sets in at
$\mu_I=m_\pi$, where the triangle is flat: the apex lies on the base, and
$k=0$.  Beyond the onset the apex rises and the base shrinks.

Figure~\ref{fig:state}d follows two angles of the triangle.  The modulus of
\eqref{eq:tridict}, $k^{2}=\cos^{2}(\theta/2)$, belongs to the angle $\theta$
at the apex; it runs from $0$ towards $1$, and at $\mu_I=4m_\pi$ it is
$0.9985$.  The second curve is $\sin^{2}(\theta_-/2)$, where $\theta_-$ is the
angle at the end $+M$ of the base, between it and the side $\ell_-$:
$\cos\theta_-=(M-\nu)/\ell_-$.  Each curve passes $\tfrac12$ where its angle is
a right angle.  For the apex that is $\nu^{2}+\Delta^{2}=M^{2}$, at
$\mu_I=1.17\,m_\pi$.  For the base it is $\nu=M$, at $\mu_I=1.67\,m_\pi$.
There the apex stands above the end $+M$ of the base, the minimum of the lower
level leaves $p=0$, and from there on the gap of the spectrum is the height
$\Delta$ of the triangle and no longer its side $\ell_-$
(section~\ref{sec:model}).  By this change in the minimum of the dispersion
the crossover from a Bose--Einstein condensate of pions to BCS pairing has
been located in this model~\cite{SunHeZhuang2007}; see also the
review~\cite{HeMaoZhuang2013}.  Nothing in the closed form is special at that
point: none of the three points \eqref{eq:abc} meet there, and the curves of
figure~\ref{fig:state} pass through it smoothly.  The two right angles are
different points.  They coincide only in the nonrelativistic limit, where
$k^{2}=\tfrac12$ marks the change of sign of the chemical potential of
ref.~\cite{MariniPistolesiStrinati1998} (appendix~\ref{app:checks}).

With the cutoff kept the state is a different one.  It stays close to the
first near the onset and then moves away: at $\mu_I=4m_\pi$ its $\Delta$ is
$342$~MeV against $315$, and its density is lower by $11$ per cent.  The two
are the same integral with the same vacuum, treated differently at one end of
the path.  Their difference is the tail of order $\Lambda^{-2}$ of
\eqref{eq:asymp}, and it acts twice: directly, and through the three numbers,
which it moves in the fit to the vacuum.  It is not small because
$\nu/\Lambda$ is not: at $\mu_I=4m_\pi$ it is $0.4$.  The reduction does not
decide which of the two is the model.  What it does is to make the question
precise: the tail is explicit, made of the boundary term and the logarithm
taken at the cutoff, each minus its limit, and of the three bounded integrals
of \eqref{eq:identlog} between $\varphi_\Lambda$ and $\varphi_1$.

\section{Conclusions}\label{sec:conclusions}
We have obtained a closed form for the zero-temperature mean-field
thermodynamic potential of relativistic fermions with a mass, a chemical
potential and a gap.  The potential contains a divergent momentum integral of
the sum of the two quasiparticle levels.  The Hermite reduction evaluates this
integral exactly at an arbitrary cutoff, reducing it to an algebraic boundary
term and three incomplete elliptic integrals, one of each kind.  Their modulus,
parameter and amplitude at the end of the path are fixed by the triangle
formed by the two levels at zero momentum.  All power divergences reside in the
boundary term.  The only remaining divergence is logarithmic, with the
coefficient $2A_3$ fixed by the residue at the puncture where the path of
integration ends.  A change of parameter in the elliptic integral of the third
kind isolates this logarithm explicitly at every cutoff.

A regularization is then a prescription for the end of the path.  For sharp and
smooth state-independent cutoffs acting on the momentum alone, the regulator
dependence of the finite part is confined, as the cutoff is removed, to the span
of $1$, $\rho^{2}$ and $A_3$.  The corresponding shifts enter an additive
constant, the coupling and the scale of the logarithm.  Once the vacuum quark
mass $M_0$, the pion decay constant $f_\pi$ and the pion mass $m_\pi$ fix the
coupling, the scale of the logarithm and the current quark mass, all such
schemes give the same potential up to an additive constant.  Everything outside
these three conventional terms, including the part of the boundary constant
outside the span, is a prediction of the model.  The medium separation scheme,
as implemented in ref.~\cite{Avancini2019}, reproduces this potential up to a
constant.  A cutoff applied separately to the two levels does not: it generates
an isospin density below the onset of condensation and violates the Silver
Blaze property.  Keeping a momentum cutoff finite retains explicit terms of
order $\Lambda^{-2}$ and therefore defines a different model rather than
another regularization of the same one.

Along the isospin axis, the closed form gives the local minimum that continues
the vacuum without numerical momentum integration, across both the onset of
pion condensation and the point at which the minimum of the lower level leaves
zero momentum.  The derivation assumes zero temperature, equal masses of the
paired fermions and a uniform, momentum-independent gap.  Beyond them, as in a
two-colour model with split masses and a gap that carries a momentum, the
quasiparticle energies are not known in closed form and are found
numerically~\cite{AndersenBrauner2010}.  Within them the result depends only
on the two-level spectrum.  It therefore applies unchanged to the
mean-field NJL potential of two-colour QCD~\cite{BraunerFukushimaHidaka2009}.
It also describes paired quarks with equal chemical potentials in a two-flavour
colour superconductor, mass included, with the unpaired quarks contributing an
additional free-gas term~\cite{Shovkovy2005}.  The same integral occurs at
fixed mass in the relativistic BCS--BEC crossover, where the logarithm survives
when the coupling is traded for a scattering length and must still be
regulated~\cite{Brauner2008}.  The elliptic structure thus belongs to the
two-level spectrum of relativistic pairing, not to pion condensation alone.

The reduction does not depend on there being a single condensate.  We call a
condensate and a chemical potential a pair when the charge of the potential
anticommutes with the term of the condensate in the mean-field Hamiltonian and
commutes with all its other terms.  The pion condensate and the isospin
chemical potential form such a pair, and the quark mass, which has no
potential of its own, stays outside.  With $n$ pairs the quasiparticle levels
are the sums, with all signs, of the $n+1$ singular values of a matrix built
from the momentum, the condensates and the potentials, and the integrand is an
algebraic function on a curve of genus $n$, or $n-1$ if no mass is left
outside the pairs.  As long as no level crosses zero, the same reduction goes
through with the same boundary term.  All power divergences remain in it, and
the only other divergence is again a logarithm with the coefficient $2A_3$, to
which each pair contributes its own $\nu_j^{2}\Delta_j^{2}$.  For cutoffs
acting on the momentum alone the regulator dependence is therefore again
confined to the span of $1$, $\rho^{2}$ and $A_3$.  The closed form in
Legendre's integrals is the case of genus one.  Besides the case treated here,
it covers the chiral isospin chemical
potential~\cite{KhunjuaKlimenkoZhokhov2019}, which pairs with the quark mass
and leaves the curve of genus one.  A further pair, such as a diquark
condensate with the quark-number chemical potential in two-colour
QCD~\cite{KhunjuaKlimenkoZhokhov2020}, raises the genus to two.  The remaining
integrals are then hyperelliptic and, for parameters in general position, do
not reduce to elliptic ones.

\acknowledgments
\begin{sloppypar}
This work began in 2009, and I am deeply grateful to my supervisors at the
time, Prof.~Zhukovsky and Prof.~Klimenko.  I owe particular thanks
to A.~Akhmeteli, whose help at an early stage was invaluable and who first pointed me
towards elliptic integrals.  I also thank Academician~Kvinikhidze and Dr.~Zhokhov
for valuable discussions.
\end{sloppypar}

This work was supported by the Shota Rustaveli National Science Foundation of
Georgia (SRNSFG) under grant no.~FR-25-7544, ``Nonperturbative Analysis of the
QCD Critical Point: Regularization problems in the Effective Models and
Renormalization Group Techniques''.

During the preparation of the manuscript, Claude (Anthropic, Claude Opus 5.5)
was used to edit the English text and helped to write and assemble the
supplementary material, together with the script that generates and checks it.
The author reviewed and revised
all AI-assisted text and takes full responsibility for the scientific content
and final wording. 

\appendix

\section{From the integral to the identity}\label{app:identity}
By \eqref{eq:p2x}
$p^{2}$ is a rational function of $R$, and its partial fractions are
\begin{equation}\label{eq:pfrac}
  p^{2}=\frac R4-\rho^{2}-\frac{4\,\nu^{2}\Delta^{2}}{R-4\nu^{2}} ,
\end{equation}
where the residue at $R=b$ is $-4\kappa$, with $\kappa=\nu^{2}\Delta^{2}$:
the medium enters through the strength of one pole.  With
$dp=\tfrac12(p^{2})'\,dR/p$ and \eqref{eq:psqrtR}, the measure is
\begin{equation}\label{eq:Qx}
  \begin{split}
  p^{2}\sqrt R\,dp&=Q(R)\,\frac{dR}{Y},\qquad
  Q=\frac{R(R-a)(R-c)}{4}\,\frac{d(p^{2})}{dR}\\
  &=\frac{R(R-a)(R-c)}{16}+\kappa\,\frac{R(R-a)(R-c)}{(R-b)^{2}} ,
  \end{split}
\end{equation}
since $d(p^{2})/dR=\tfrac14+4\kappa/(R-b)^{2}$ by \eqref{eq:pfrac}.  The
first term of $Q$ is what a free particle would give, and the second is the
medium.  Together they are a cubic polynomial plus a double and a simple pole
at $b$.

An integration by parts subtracts a derivative.  With $P_4\equiv Y^{2}$ and
an algebraic factor $f$, the product $Y\,d(fY)/dR=f'P_4+\tfrac12fP_4'$ is a
rational function of $R$; subtracting a multiple of it from $Q$ costs the same multiple of $fY$
at the ends of the path, and the multiple is the ratio of the coefficients of
the piece to be removed in $Q$ and in $f'P_4+\tfrac12fP_4'$.  For
$f=(R-b)^{-m}$ the numerator of the derivative is $(R-b)P_4'-2mP_4$, which
vanishes at the branch point $R=b$, where $P_4=0$; so the derivative has a
pole of order $m$, not $m+1$.  The move with $m=1$
therefore removes $R^{2}$ at the price of a simple pole at $b$, and the move
with $m=2$ lowers the double pole to a simple one.  The simple pole is kept,
because the normal form below turns it into an integral of the second kind.
Three moves, in this order:
\begin{equation}\label{eq:threemoves}
  {\renewcommand{\arraystretch}{2.0}
  \begin{array}{lll}
  f=1, & \text{removes }R^{3}: & \dfrac1{32},\\
  f=(R-b)^{-1}, & \text{removes }R^{2}: & s_1=\dfrac{3b-a-c}{64}=\dfrac{2\nu^{2}-\rho^{2}}{16},\\
  f=(R-b)^{-2}, & \text{removes }(R-b)^{-2}: &
     s_2=-\dfrac{(a-b)(b-c)}{24}=-\dfrac{2}{3}\,\nu^{2}\Delta^{2}.
  \end{array}}
\end{equation}
Their multiples are the coefficients of the boundary term \eqref{eq:Wdef},
and what they leave is the remainder of \eqref{eq:remx}, with
\begin{equation}\label{eq:Acoef}
  A_1 = \frac{(a - b)}{384}\,\bigl[-3 a b + 15 b^{2} + 8 a c - 33 b c + 16 c^{2}\bigr],
\end{equation}
\begin{equation}\label{eq:Ccoef}
  A_2 = \frac{(a - b)}{384}\,\bigl[-13 a b + 15 b^{2} + 8 a c - 13 b c\bigr],
\end{equation}
\begin{equation}\label{eq:Bcoef}
  A_3 = -\frac{1}{128}\,\bigl[a^{2} - 6 a b + 5 b^{2} + 6 a c - 6 b c + c^{2}\bigr].
\end{equation}
Once $dW=(f'P_4+\tfrac12fP_4')\,dR/Y$ is written out for the three factors,
\eqref{eq:remx} is an equality between rational functions of $R$, and it is
checked by multiplying out.  In the variables of the model $A_3$ is
\eqref{eq:residue}.  Integrated along the path, \eqref{eq:remx} reads
\begin{equation}\label{eq:lowerx}
  \begin{split}
  \int_a^{R_\Lambda}Q\,\frac{dR}{Y}
  &=W(R_\Lambda)-W(a)\\
  &\quad+\int_a^{R_\Lambda}\Bigl[A_1+A_2\frac{R-c}{R-b}+A_3(R-b)\Bigr]\frac{dR}{Y},
  \qquad W(a)=0 :
  \end{split}
\end{equation}
the lower limit is the branch point $a$, where $Y$ vanishes and the bracket of
\eqref{eq:Wdef} is finite.

The three integrals left are brought to Legendre's normal form by the
substitution 258.00 of Byrd and Friedman~\cite{ByrdFriedman1971}, written for
four real points $a>b>c>d$ and a path $R\ge a$; here $d=0$:
\begin{equation}\label{eq:legsub}
  \sn^{2}u=\frac{b\,(R-a)}{a\,(R-b)},\qquad
  k^{2}=\frac{a\,(b-c)}{b\,(a-c)},\qquad
  \alpha^{2}=\frac{a}{b},\qquad
  g=\frac{2}{\sqrt{b\,(a-c)}} ,
\end{equation}
with $0<k^{2}<1$ and $\alpha^{2}>1$ by the order of the points.  Under it
$R=a$ is $u=0$, and with $D\equiv1-\alpha^{2}\sn^{2}u$ one has
$R-b=(a-b)/D$, $(R-c)/(R-b)=[(a-c)/(a-b)]\,\dn^{2}u$ and $dR/Y=g\,du$.  With
$\sn u=\sin\varphi$ the three integrals are therefore
\begin{equation}\label{eq:legdict}
  \begin{gathered}
  \int_a^{R}\frac{dR}{Y}=g\,\EllF(\varphi,k),\qquad
  \int_a^{R}\frac{R-c}{R-b}\,\frac{dR}{Y}=g\,\frac{a-c}{a-b}\,\EllE(\varphi,k),\\
  \int_a^{R}(R-b)\,\frac{dR}{Y}=g\,(a-b)\,\Pi(\varphi;\alpha^{2},k).
  \end{gathered}
\end{equation}
With \eqref{eq:lowerx} and $\int_0^{\Lambda}p^{2}\sqrt R\,dp=\int_a^{R_\Lambda}Q\,dR/Y$,
this is \eqref{eq:identity}.

\section{The upper limit and the finite part}\label{app:finite}
At $p=\Lambda$ the path ends at $R_\Lambda$, and both it and the root $Y$ there
are explicit, by \eqref{eq:Rdef} and \eqref{eq:psqrtR}:
\begin{equation}\label{eq:RLam}
  R_\Lambda=\bigl[E^{+}(\Lambda)+E^{-}(\Lambda)\bigr]^{2}=4\Lambda^{2}+4\rho^{2}+O(\Lambda^{-2}),
  \qquad
  Y_\Lambda=2\Lambda\,(R_\Lambda-b)\sqrt{R_\Lambda} .
\end{equation}
The boundary term there is a Laurent series in $\Lambda$ with three terms that
do not vanish:
\begin{equation}\label{eq:WLam}
  W(R_\Lambda)=\frac{\Lambda^{4}}{2}+\frac{\rho^{2}\Lambda^{2}}{2}+\mathcal C_W+O(\Lambda^{-2}),
  \qquad
  \mathcal C_W=\frac{\nu^{2}\Delta^{2}}{12}+\frac{\rho^{4}}{16} .
\end{equation}
The proof is a count of degrees.  In the variable $\sqrt{R_\Lambda}$,
$\Lambda^{2}$ is rational by \eqref{eq:p2x}, and so is $W^{2}$; both $W^{2}$
and the square of the right side have degree eight at infinity, and their
difference has degree two, so the two square roots differ by
$O(1/R_\Lambda)=O(\Lambda^{-2})$.

Of the three integrals the first two tend to $\EllF(\varphi_1,k)$ and
$\EllE(\varphi_1,k)$ as $O(\Lambda^{-2})$, since
$\sin^{2}\varphi_\Lambda-\sin^{2}\varphi_1=O(1/R_\Lambda)$.  The third
diverges, at a known rate.  Along the path $D=(a-b)/(R-b)$, so that at the
cutoff $D_\Lambda=(a-b)/(4\Lambda^{2})+O(\Lambda^{-4})$, and
\begin{equation}\label{eq:Pilog}
  \Pi(\varphi_\Lambda;\alpha^{2},k)=-L\,\ln D_\Lambda+O(1),\qquad
  L=\frac{1}{g\,(a-b)} ,
\end{equation}
because $d[\Pi+L\ln D]/du=(1-2L\alpha^{2}\sn u\cn u\dn u)/D$, and both the
numerator and $D$ vanish simply at the end of the path $u_1$.  There
$\sn^{2}u_1=1/\alpha^{2}$, so that
$(\sn u_1\cn u_1\dn u_1)^{2}=(\alpha^{2}-1)(\alpha^{2}-k^{2})/\alpha^{6}$, and
$4L^{2}\alpha^{4}$ times this is one, an identity in the three points.  Hence
\begin{equation}\label{eq:lograte}
  g\,(a-b)\,A_3\,\Pi(\varphi_\Lambda;\alpha^{2},k)
  =-A_3\ln D_\Lambda+O(1)=2A_3\ln\Lambda+O(1):
\end{equation}
the logarithm of the cutoff, with twice the residue \eqref{eq:residue} as its
coefficient.  Its finite remainder needs one relation from the tables, the
change of parameter of the third kind:
\tabbox{DLMF 19.7.8 with 19.2.17 and 19.2.19 --- change of parameter}{%
Let $c=\csc^{2}\varphi\neq\alpha^{2}$ and $\alpha^{2}\omega^{2}=k^{2}$.  Then
\[
  \Pi(\varphi;\alpha^{2},k)+\Pi(\varphi;\omega^{2},k)
  =\EllF(\varphi,k)+\sqrt{c}\,R_C\bigl((c-1)(c-k^{2}),\,(c-\alpha^{2})(c-\omega^{2})\bigr),
\]
with Carlson's degenerate integral
$R_C(x,y)=\tfrac12\int_0^{\infty}dt\,(t+x)^{-1/2}(t+y)^{-1}$, which for
$0<y<x$ is $R_C(x,y)=\operatorname{artanh}\sqrt{(x-y)/x}\,\big/\sqrt{x-y}$ and
hence, as $y\to0^{+}$, $R_C(x,y)=\dfrac{1}{2\sqrt x}\ln\dfrac{4x}{y}+O(y\ln y)$.
The square root on $c$ is easy to lose in transcription, and the relation
fails without it.  (In this box $c$ is the tables' letter for $\csc^{2}\varphi$,
not the point of \eqref{eq:abc}.)}
Along the path $\csc^{2}\varphi$ is a rational function of $R$ by
\eqref{eq:legsub}, and
\begin{equation}\label{eq:degen}
  \csc^{2}\varphi-\alpha^{2}=\frac{\alpha^{2}\,(a-b)}{R-a}:
\end{equation}
the second argument of $R_C$ vanishes like $1/R_\Lambda$ as the cutoff goes to
infinity: the change of parameter degenerates at the one point where the pole
meets the end of the path, and that is where the logarithm comes from.  At the
endpoint $\csc^{2}\varphi=\alpha^{2}$ the first argument is
$x_1=(\alpha^{2}-1)(\alpha^{2}-k^{2})$, and $\alpha/(2\sqrt{x_1})=L$, the same
$L$ as in \eqref{eq:Pilog}.  With
$\csc^{2}\varphi-\alpha^{2}=\alpha^{2}D_\Lambda\,[1+O(\Lambda^{-2})]$ the
elementary term is
$-L\ln D_\Lambda+L\ln\bigl[4x_1/(\alpha^{2}(\alpha^{2}-\omega^{2}))\bigr]+o(1)$,
and its constant simplifies in the three points, and then in the triangle:
\begin{equation}\label{eq:logconst}
  \begin{gathered}
  \frac{16\,(\alpha^{2}-1)(\alpha^{2}-k^{2})}{(a-b)\,\alpha^{2}(\alpha^{2}-\omega^{2})}
  =\frac{16}{a+b-c}=\frac{4}{\ell_+\ell_-+\nu^{2}},\\
  \omega^{2}=\frac{b-c}{a-c},\qquad \sin^{2}\varphi_1=\frac{b}{a},
  \end{gathered}
\end{equation}
where the factor $a-b$ on the left is the one that $D_\Lambda$ brings in.
Put into \eqref{eq:identity}, with
$-A_3\ln D_\Lambda=2A_3\ln\Lambda+A_3\ln[4/(a-b)]+O(\Lambda^{-2})$ and with
\eqref{eq:WLam}, this gives the finite part \eqref{eq:Cfin}.

The change of parameter can be made at the cutoff itself, not only in the
limit.  Along the path it reads
$\Pi(\varphi;\alpha^{2},k)+\Pi(\varphi;\omega^{2},k)=\EllF(\varphi,k)+2L\operatorname{artanh}t$,
with $t=Y/[R\,(R-c)]$; both sides vanish at $R=a$, and their derivatives in
$R$ agree.  Put into \eqref{eq:identity}, with $g\,(a-b)\,L=1$, it gives
\begin{equation}\label{eq:identlog}
  \begin{split}
  \int_0^{\Lambda}p^{2}\,\bigl[E^{+}+E^{-}\bigr]\,dp
  &=W(R_\Lambda)+2A_3\operatorname{artanh}t_\Lambda\\
  &\quad+g\Bigl[(A_1+(a-b)A_3)\,\EllF(\varphi_\Lambda,k)+A_2\,\frac{a-c}{a-b}\,\EllE(\varphi_\Lambda,k)\\
  &\qquad\quad-(a-b)\,A_3\,\Pi(\varphi_\Lambda;\omega^{2},k)\Bigr],\qquad
  t_\Lambda=\frac{Y_\Lambda}{R_\Lambda\,(R_\Lambda-c)} ,
  \end{split}
\end{equation}
exact at every $\Lambda$.  The powers of the cutoff are in the boundary term,
the logarithm is the one elementary function
$2A_3\operatorname{artanh}t_\Lambda$, and the bracket is bounded.  Since
$1-t_\Lambda^{2}=(a+b-c)/R_\Lambda+O(R_\Lambda^{-2})$, the logarithm is
$A_3\ln[16\Lambda^{2}/(a+b-c)]+O(\Lambda^{-2})$, and \eqref{eq:asymp} and
\eqref{eq:Cfin} follow again.  The remainder $O(\Lambda^{-2})$ of
\eqref{eq:asymp} is thus explicit: it is the sum of the terms of
\eqref{eq:identlog}, each minus its limit.

\section{The medium separation scheme}\label{app:mss}
The integrals of the two carriers of
section~\ref{sec:removed} to the cutoff are elementary: the first is the free
case of section~\ref{sec:parts} at the mass $\rho$, and the second is
\begin{equation}\label{eq:carriers}
  \int_0^{\Lambda}\frac{p^{2}\,dp}{E_\rho^{3}}
  =\operatorname{arsinh}\frac{\Lambda}{\rho}-\frac{\Lambda}{\sqrt{\Lambda^{2}+\rho^{2}}}
  =\ln\frac{2\Lambda}{\rho}-1+O(\Lambda^{-2}) ,
\end{equation}
and in the logarithm of \eqref{eq:Fclass} the cutoff is replaced by
$\lambda$.

The scheme has been applied to this model, and its potential is written out
in equation~(2.28) of ref.~\cite{Avancini2019}.  In our notation (their
$\mu_I$ is our $\nu$, their $\sigma$ is $M-m_0$) it reads
\begin{equation}\label{eq:mss}
  \begin{split}
  \Omega_{\rm MSS}&=\frac{(M-m_0)^{2}+\Delta^{2}}{4G}
  -\frac{N_c}{\pi^{2}}\Bigl\{\tilde M\,I_{\rm q}
  -\frac{\tilde M^{2}-4\kappa}{4}\,I_{\ln}+\mathcal R_s\Bigr\},\\
  \mathcal R_s&=\int_0^{\infty}p^{2}\Bigl[E^{+}+E^{-}-2E_s-\frac{\tilde M}{E_s}
  +\frac{\tilde M^{2}-4\kappa}{4E_s^{3}}\Bigr]dp,\\
  \tilde M&=\rho^{2}-M_s^{2},\qquad E_s=\sqrt{p^{2}+M_s^{2}} ,
  \end{split}
\end{equation}
with $\kappa=\nu^{2}\Delta^{2}$.  Here $M_s$ is the mass at which the
subtraction is made.  In the reference it is the vacuum mass; it has its own
letter here because the identity below does not use that.  $I_{\rm q}$ and
$I_{\ln}$ are the two divergent vacuum integrals, $2\pi^{2}$ times the
$I_{\rm quad}$ and $I_{\rm log}$ of the reference.  The scheme integrates the
remainder $\mathcal R_s$ over all momenta and cuts the two vacuum integrals at
$\Lambda$:
\begin{equation}\label{eq:mssI}
  \begin{split}
  I_{\rm q}&=\int_0^{\Lambda}\frac{p^{2}\,dp}{E_s}
  =\frac12\Bigl[\Lambda E_{s\Lambda}-M_s^{2}\operatorname{arsinh}\frac{\Lambda}{M_s}\Bigr],\\
  I_{\ln}&=\int_0^{\Lambda}\frac{p^{2}\,dp}{E_s^{3}}
  =\operatorname{arsinh}\frac{\Lambda}{M_s}-\frac{\Lambda}{E_{s\Lambda}} ,
  \end{split}
\end{equation}
where $E_{s\Lambda}=\sqrt{\Lambda^{2}+M_s^{2}}$.  The cutoff enters
\eqref{eq:mss} through these two numbers and nowhere else.

The three subtracted terms have the three leading terms of \eqref{eq:puncture},
$2p+\rho^{2}/p+2A_3/p^{3}$: they are the two carriers expanded about the mass
$M_s$ instead of taken at $\rho$.  The integrand of $\mathcal R_s$ therefore
falls as $p^{-3}$, and the remainder converges.  The integrals of the three
subtracted terms to a cutoff are elementary, the free case at the mass $M_s$
and \eqref{eq:mssI}; subtracted from \eqref{eq:asymp}, they cancel its two
powers of the cutoff and its logarithm, and what is left is
\begin{equation}\label{eq:mssR}
  \mathcal R_s=\mathcal C+2A_3\ln\frac{M_s}{2}-\frac{M_s^{4}}{16}
  -\frac{\tilde M M_s^{2}}{4}-\frac{\tilde M^{2}-4\kappa}{4} .
\end{equation}
Inserted into \eqref{eq:mss}, with
$\tfrac14(\tilde M^{2}-4\kappa)=-2A_3-\tfrac12\rho^{2}M_s^{2}+\tfrac14M_s^{4}$,
this gives a potential made of $\mathcal C$, $A_3$, $\rho^{2}$, $M$ and a
constant, and that is \eqref{eq:mssmap}.  The identity does not use
\eqref{eq:mssI}: whatever values the two vacuum integrals are given, they enter
only as $A'$ and $\lambda$.  With the sharp cutoff the two inverse hyperbolic
sines combine, and
\begin{equation}\label{eq:mssclosed}
  \lambda=\frac{\Lambda+E_{s\Lambda}}{2}\,e^{\,1-\Lambda/E_{s\Lambda}},\qquad
  A'=\frac1{2G}-\frac{N_c}{\pi^{2}}\Bigl[\frac{\Lambda^{3}}{E_{s\Lambda}}+\frac{M_s^{2}}{2}\Bigr] .
\end{equation}
For $\Lambda\gg M_s$ these are $\lambda=\Lambda\,[1+\tfrac34M_s^{2}/\Lambda^{2}+\dots]$
and $A'=1/(2G)-N_c\Lambda^{2}/\pi^{2}+O(\Lambda^{-2})$, the $A'$ of
\eqref{eq:Omclass}.  At a finite cutoff $\lambda>\Lambda$, because
$E_{s\Lambda}>\Lambda$ makes both factors of \eqref{eq:mssclosed} larger: the
scale of the logarithm lies above the cutoff it came from.

Let the subtraction be made
at the vacuum mass of the scheme, $M_s=M_0$.  In the onset condition of
section~\ref{sec:removed}, the coefficient of $\Delta^{2}$ in the integrand
leaves, after the two carriers, the convergent remainder $\nu^{4}J(\nu,M)$,
with $J(\nu,M)=\int_0^{\infty}p^{2}dp/[E_p^{3}(E_p^{2}-\nu^{2})]$.  With
\eqref{eq:mssmap} the three vacuum conditions
\eqref{eq:gapvac}--\eqref{eq:thouless} then read
\begin{equation}\label{eq:mssvac}
  \begin{gathered}
  \frac{M_0-m_0}{2G}=\frac{2N_c}{\pi^{2}}\,M_0\,I_{\rm q},\qquad
  f_\pi^{2}=\frac{N_cM_0^{2}}{2\pi^{2}}\,I_{\ln},\\
  \frac1{4G}=\frac{N_c}{\pi^{2}}\Bigl[I_{\rm q}+\nu^{2}I_{\ln}+\nu^{4}J(\nu,M_0)\Bigr]_{\nu=m_\pi/2} .
  \end{gathered}
\end{equation}
The first is the gap equation of the model with the sharp cutoff at
$\nu=\Delta=0$: the subtraction mass is the vacuum mass of the scheme exactly
when it is the vacuum mass of that model.  The second is the integral of
\eqref{eq:fpivac} cut at $\Lambda$.  The third is the onset condition of the
model with the sharp cutoff in all but its last term: there $J$ is cut at
$\Lambda$ as well, here it is not.  So the scheme is the potential
\eqref{eq:Omclass} with the vacuum mass and the decay constant of the model
with the sharp cutoff at the same $G$, $m_0$ and $\Lambda$, and with a slightly
earlier onset, by the tail of $J$ beyond the cutoff: by $0.4$~MeV in $\mu_I$
for the model of section~\ref{sec:axis}.

\section{Derivatives of the closed form}\label{app:derivatives}
The potential depends on the state through the chain
$(M,\Delta,\nu)\to(\ell_+,\ell_-)\to(a,b,c)\to\{k^{2},\omega^{2},\sin^{2}\varphi_1,g,A_1,A_2,A_3,\mathcal C_W\}\to\mathcal C$,
every link of which but the last is algebraic.  The points:
\begin{equation}\label{eq:dpoints}
  \begin{gathered}
  \frac{\partial a}{\partial x}=2\,(\ell_++\ell_-)\Bigl(\frac{\partial\ell_+}{\partial x}+\frac{\partial\ell_-}{\partial x}\Bigr),\qquad
  \frac{\partial c}{\partial x}=2\,(\ell_+-\ell_-)\Bigl(\frac{\partial\ell_+}{\partial x}-\frac{\partial\ell_-}{\partial x}\Bigr),\\
  \frac{\partial b}{\partial\nu}=8\nu,\\
  {\renewcommand{\arraystretch}{2.4}
  \begin{array}{l|ccc}
   x & M & \Delta & \nu\\ \hline
   \partial\ell_+/\partial x & \dfrac{M+\nu}{\ell_+} & \dfrac{\Delta}{\ell_+} & \dfrac{M+\nu}{\ell_+}\\
   \partial\ell_-/\partial x & \dfrac{M-\nu}{\ell_-} & \dfrac{\Delta}{\ell_-} & -\dfrac{M-\nu}{\ell_-}
  \end{array}}
  \end{gathered}
\end{equation}
The coefficients $A_1$, $A_2$, $A_3$, $\mathcal C_W$ are polynomials in the
points; the other Legendre data and the argument of the logarithm have the
logarithmic derivatives
\begin{equation}\label{eq:dleg}
  {\renewcommand{\arraystretch}{2.5}
  \begin{array}{l|ccc}
   \partial_x\ln(\cdot) & x=a & x=b & x=c\\ \hline
   k^{2}=\dfrac{a\,(b-c)}{b\,(a-c)} & \dfrac1{a}-\dfrac1{a-c} & \dfrac1{b-c}-\dfrac1{b} & \dfrac1{a-c}-\dfrac1{b-c}\\
   \omega^{2}=\dfrac{b-c}{a-c} & -\dfrac1{a-c} & \dfrac1{b-c} & \dfrac1{a-c}-\dfrac1{b-c}\\
   \sin^{2}\varphi_1=\dfrac{b}{a} & -\dfrac1{a} & \dfrac1{b} & 0\\
   g=\dfrac{2}{\sqrt{b\,(a-c)}} & -\dfrac1{2(a-c)} & -\dfrac1{2b} & \dfrac1{2(a-c)}\\
   a+b-c & \dfrac1{a+b-c} & \dfrac1{a+b-c} & -\dfrac1{a+b-c}
  \end{array}}
\end{equation}
The derivatives of $\EllF$, $\EllE$ and $\Pi$ with respect to the amplitude are
their integrands, with $\Delta_\varphi=\sqrt{1-k^{2}\sin^{2}\varphi}$; with
respect to the parameters,
\begin{equation}\label{eq:dFEP}
  \begin{split}
  \frac{\partial \EllF}{\partial k}&=\frac{\EllE-k'^{2}\EllF}{k\,k'^{2}}-\frac{k\sin\varphi\cos\varphi}{k'^{2}\,\Delta_\varphi},\qquad
  \frac{\partial \EllE}{\partial k}=\frac{\EllE-\EllF}{k},\\
  \frac{\partial\Pi}{\partial k}&=\frac{k}{k'^{2}(k^{2}-\omega^{2})}
   \Bigl[\EllE-k'^{2}\Pi-\frac{k^{2}\sin\varphi\cos\varphi}{\Delta_\varphi}\Bigr],\\
  \frac{\partial\Pi}{\partial\omega^{2}}&=\frac{(\omega^{2}-k^{2})\EllF+(k^{2}-\omega^{4})\Pi-\omega^{2}\EllE}{2\omega^{2}(\omega^{2}-1)(\omega^{2}-k^{2})}
   +\frac{\omega^{2}\sin\varphi\cos\varphi\,\Delta_\varphi}{2(\omega^{2}-1)(\omega^{2}-k^{2})(1-\omega^{2}\sin^{2}\varphi)},
  \end{split}
\end{equation}
all at the same $(\varphi,k,\omega^{2})$, with $k'^{2}=1-k^{2}$.  Each line is
an identity between integrands plus $d/d\varphi$ of an algebraic term that
vanishes at $\varphi=0$.  Every derivative of $\mathcal C$, of any order, is
therefore again algebraic in the points times $\EllF$, $\EllE$ and $\Pi$, plus a
polynomial times the logarithm of \eqref{eq:Cfin}, plus algebraic terms.

\section{Checks against known limits}\label{app:checks}
At $\Delta=0$ and $\nu<M$ the two middle points coincide, $b=c$, so that
$k=\omega=0$ and $A_1+A_2=0$; the elliptic terms drop out, and with $a=4M^{2}$
\begin{equation}\label{eq:delta0}
  \mathcal C\big|_{\Delta=0}=\mathcal C_W\big|_{\Delta=0}+A_3\ln\frac{4}{M^{2}}
  =\frac{M^{4}}{16}-\frac{M^{4}}{8}\ln\frac{4}{M^{2}} ,
\end{equation}
the finite part of the free case at the mass $M$, independent of $\nu$.  At
$\Delta=0$ and $\nu>M$ the levels are $E_p+\nu$ and $|E_p-\nu|$, free quarks
with a Fermi sphere of radius $p_F=\sqrt{\nu^{2}-M^{2}}$, and the integral is
elementary again:
\begin{equation}\label{eq:delta0F}
  \mathcal C\big|_{\Delta=0,\,\nu>M}=\frac{M^{4}}{16}-\frac{M^{4}}{8}\ln\frac{4}{M^{2}}
  +\frac{\nu p_F^{3}}{6}-\frac{M^{2}\nu p_F}{4}+\frac{M^{4}}{4}\ln\frac{\nu+p_F}{M} ,
\end{equation}
the free case plus the Fermi sea.  Here the two upper points coincide, $a=b$,
so that $k=1$ and $\varphi_1=\pi/2$, and the closed form reaches
\eqref{eq:delta0F} as a limit and not by substitution: the first correction is
$\nu p_F\,\Delta^{2}\ln(1/\Delta)$, which is not analytic in $\Delta$.

At $M=0$ the two lowest points coincide, $c=0$, and $k=1$.  The three integrands
are then rational in $\sin\varphi$, the relation $A_1=bA_3$ cancels every
$\operatorname{artanh}(\sin\varphi_1)$, and with $a=4(\nu^{2}+\Delta^{2})$,
$b=4\nu^{2}$
\begin{equation}\label{eq:M0}
  \mathcal C\big|_{M=0}=\frac{\nu^{4}}{6}-\nu^{2}\Delta^{2}+\frac{\Delta^{4}}{16}
  +\frac{4\nu^{2}\Delta^{2}-\Delta^{4}}{8}\,\ln\frac{4}{\Delta^{2}} .
\end{equation}
The same integral done directly is elementary and gives the same finite part.
In $\Omega$ the first term is $-N_c\nu^{4}/(6\pi^{2})$, the pressure of free
massless quarks of two flavours; it is not in $\mathcal C_W$, and it is
assembled by the elliptic terms as they degenerate.

Let $M\to\infty$ with $\mu\equiv\nu-M$ and $\Delta$ fixed.  The level $E^{-}$
becomes $\sqrt{(p^{2}/2M-\mu)^{2}+\Delta^{2}}$, the spectrum of a
nonrelativistic superfluid with the chemical potential $\mu$, and $E^{+}$ goes
to $2M$.  In this limit
\begin{equation}\label{eq:nonrel}
  \begin{gathered}
  k^{2},\ \omega^{2}\ \longrightarrow\ k_{\rm nr}^{2}=\frac12\Bigl(1+\frac{x_0}{\sqrt{1+x_0^{2}}}\Bigr),
  \qquad
  \cos^{2}\varphi_1=\frac{\ell_--\mu}{M}+O(M^{-2}),\qquad x_0=\frac{\mu}{\Delta},\\
  -\frac{\partial^{2}\mathcal C}{\partial M\,\partial\nu}
  =\int_0^{\infty}\frac{M\Delta^{2}}{E_p}\Bigl[\frac{1}{(E^{-})^{3}}-\frac{1}{(E^{+})^{3}}\Bigr]p^{2}\,dp
  \ \longrightarrow\ (2M)^{3/2}\,\Delta^{1/2}\,I_5(x_0) .
  \end{gathered}
\end{equation}
The amplitude tends to $\pi/2$, the integrals become complete, and at
$\omega^{2}=k^{2}$ the integral of the third kind is
$\Pi(\tfrac{\pi}{2};k^{2},k)=\EllE(k)/(1-k^{2})$.  What is left is $\EllK(k_{\rm nr})$
and $\EllE(k_{\rm nr})$, the functions and the modulus of Marini, Pistolesi and
Strinati~\cite{MariniPistolesiStrinati1998}, and
$I_5(x_0)=\int_0^{\infty}x^{2}\,[(x^{2}-x_0)^{2}+1]^{-3/2}\,dx$ is one of their
integrals.  The mixed
derivative converges by itself, since the divergent terms of \eqref{eq:asymp}
do not contain $M$ and $\nu$ together; with $p^{2}=2M\Delta\,x^{2}$ the
integrand becomes theirs.  The closed form \eqref{eq:Cfin}, differentiated,
reproduces the integral in \eqref{eq:nonrel}, and at $\Delta=1$,
$\mu=0.7$ the integral differs from its limit by $0.47$ per cent at $M=10^{2}$
and by $5\cdot10^{-5}$ at $M=10^{4}$.

\bibliographystyle{JHEP}
\bibliography{refs}
\end{document}